\documentclass[]{aa}  

\usepackage{amsmath}
\usepackage{tablefootnote}
\usepackage{bm}
\usepackage{xcolor}
\usepackage[normalem]{ulem}
\usepackage{graphicx}
\usepackage{txfonts}\usepackage[colorlinks,linkcolor=blue,citecolor=blue,linktocpage=true,breaklinks, plainpages=false,urlcolor=blue]{hyperref}
\usepackage{xcolor}

\usepackage{rotating}
\usepackage{sidecap}
\usepackage{subcaption}
\usepackage{url}
\usepackage{soul}
\hypersetup{allcolors=blue}
\usepackage{float}
\usepackage{xcolor}
\usepackage[tracking=true,kerning=true,shrink=50]{microtype}
\usepackage{lipsum}
\usepackage{txfonts}
\usepackage[export]{adjustbox} 
\bibpunct{(}{)}{;}{a}{}{,}
\DeclareUnicodeCharacter{00A0}{~}
\usepackage[us,12hr]{datetime}
\usepackage{ulem}

\usepackage{algorithm}
\usepackage{algpseudocode}

\DeclareMathSymbol{\mh}{\mathord}{operators}{`\-}

\def\Ha{\mbox{H\hspace{0.1ex}$\alpha$}}

\begin{document} 

\title{Automatic detection of Ellerman bombs using Deep Learning}

\author{
I. J. Soler Poquet\inst{1,2}
\and
C. J. D\'{i}az Baso\inst{1, 2}
\and
L. H. M. Rouppe van der Voort\inst{1,2}
\and
G. J. M. Vissers\inst{3}
}
      
\institute{
Institute of Theoretical Astrophysics,
University of Oslo, %
P.O. Box 1029 Blindern, N-0315 Oslo, Norway
\and
Rosseland Centre for Solar Physics,
University of Oslo, %
P.O. Box 1029 Blindern, N-0315 Oslo, Norway
\and
TNO, Oude Waalsdorperweg 63,
2597 AK Den Haag, The Netherlands
\\
\email{i.j.s.poquet@astro.uio.no}
}

\date{Draft: compiled on \today\ at \currenttime~UT}

\abstract
{
Ellerman bombs (EBs) are observable signatures of photospheric small-scale magnetic reconnection events. They can be seen as intensity enhancements of the $\Ha$ 6563~\AA\ line wings and as brightenings in the SDO/AIA 1600\,\AA\ and 1700\,\AA\ passbands. The reliable automatic and systematic detection of EBs would enable the study of the impact of magnetic reconnection on the Sun's dynamics.
}
{We aim to develop a method to automatically detect EBs in $\Ha$ observations from the Swedish 1-m Solar Telescope (SST) and in SDO/AIA observations using the 1600\,\AA, 1700\,\AA, 171\,\AA\ and 304\,\AA\ passbands.
}
{
We trained models based on neural networks (NNs) to perform automatic detection of EBs. Additionally, we used different types of NNs to study how different properties -- such as local spatial information, the spectral shape of each pixel, or the center-to-limb variation -- contribute to the detection of EBs.
}
{
We find that for SST observations, the NN-based models are proficient at detecting EBs. With sufficiently high spectral resolution, the spatial context is not required to detect EBs. However, as we degrade the spectral and spatial resolution, the spatial information becomes more important. Models that include both dimensions perform best. For SDO/AIA, the models struggle to reliably distinguish between EBs and bright patches of different origin. 
Permutation feature importance revealed that the $\Ha$ line wings (around $\pm1\,\AA$ from line center) are the most informative features for EB detection. For the SDO/AIA case, the 1600\,\AA\ channel is the most relevant one when used in combination with 171\,\AA\ and 304\,\AA. 
}
{
The combination of the four different SDO/AIA passbands is not informative enough to accurately classify EBs. From our analysis of a few sample SDO/AIA 1600~\AA\ and 1700~\AA\ light curves, we conclude that inclusion of the temporal variation may be a significant step towards establishing an effective EB detection method that can be applied to the extensive SDO/AIA database of observations. 
}

\keywords{Sun: photosphere -- Sun: activity -- Methods: observational --Methods: data analysis --Techniques: image processing -- Sun: magnetic fields }

\maketitle

\section{Introduction}\label{sec:introduction}

Small-scale magnetic reconnection events play a fundamental role in the dynamics of active regions. Magnetic reconnection is believed to be one of the main mechanisms for releasing energy stored in the magnetic field, heating the solar atmosphere and accelerating particles \citep{Reconnection_of_the_sun}. Since magnetic reconnection cannot be observed directly, we have to use markers to study it. A prominent marker is the so-called Ellerman bomb (EB). These phenomena were first observed by \cite{ellerman_solar_1917} and appear as compact and dynamic brightenings in active regions. Ellerman bombs are classically identified as enhancements on the wings of the \Ha\,6563\,\AA\ line, while the core remains unaffected. 

Although the main tracer of EBs is the $\Ha$ line, similar spectral signatures are observed in other hydrogen lines, such as the $\textnormal{H}\beta$ or $\textnormal{H}\varepsilon$ lines \citep{2022A&A...664A..72J, 2023A&A...677A..52K}, as well as wing enhancements in other chromospheric lines such as \ion{Ca}{ii}~8542\,\AA\ and \ion{Ca}{ii} H\,\&\,K lines \citep{socas-navarro-8564,2007A&A...473..279P, matsumoto_CaH, 2013ApJ...774...32V}. These events exhibit a flame-like morphology when seen close to the limb with sufficient spatial resolution, as first shown by \cite{2011ApJ...736...71W}. The fact that the $\Ha$ line core remains in absorption and they are not visible in the continuum indicates that EBs occur in the upper photosphere \citep{2013JPhCS.440a2007R, 2023A&A...673A..11R}. Ellerman bombs are predominantly observed near polarity inversion lines, where small magnetic field concentrations with opposite polarity meet \citep{2002ApJ...575..506G, 2007A&A...473..279P, 2011ApJ...736...71W, 2013ApJ...774...32V, 2019A&A...627A.101V, 2024A&A...683A.190R}. Some statistical studies have been performed on the general properties of EBs, though their results vary depending on the constraints used \citep[e.g.][]{2002ApJ...575..506G, 2011ApJ...736...71W, 2013ApJ...774...32V, 2013SoPh..283..307N, 2019A&A...626A...4V}. According to these studies, EBs are sub-arcsecond events, ranging from 0.1 to around 0.7~$\,\mathrm{arcsec}^2$ in size, and have lifetimes of around 3\,minutes, with some cases lasting up to about 10 min. Ellerman bomb signatures can also be observed in the mid-UV 1600$\,\AA$ and 1700\,\AA\ continua \citep{2000ApJ...544L.157Q, 2007A&A...473..279P, 2011CEAB...35..181H,2013JPhCS.440a2007R, 2019A&A...626A...4V, 2013ApJ...774...32V}. Some of these studies \citep[e.g.,][]{2013ApJ...774...32V, 2019A&A...626A...4V} have shown that the strongest EBs (when observed in the $\Ha$ line) exhibit a strong spatial and temporal correlation with the associated co-spatial middle-UV brightenings. Simulations have also been used to study the EB triggering mechanism, height of formation, and the required temperature to produce their spectral signatures \citep[e.g.,][]{2019A&A...626A..33H,2021ApJ...921...50H}.

Ellerman bombs have been detected in many studies using intensity criteria \citep[e.g.,][]{2000ApJ...544L.157Q, 2002ApJ...575..506G, 2011ApJ...736...71W, 2013ApJ...774...32V, 2019A&A...626A...4V}. In these cases, an intensity threshold is applied over one wing of $\Ha$ (or other spectral lines sensitive to EBs) or a combination of both wings, depending on the work. Pixels exceeding a given threshold are proposed as EB candidates. Temporal and spatial requirements are then used to select the final EBs. The threshold value is often related to the average intensity of the field of view (FOV). This value varies from study to study due to factors such as different instrumentation, the contrast in the data, or the observed region. A comprehensive study of this variability is detailed in Table 2 of \cite{2019A&A...626A...4V}. An important consideration regarding threshold-based methods is that they primarly rely on pixel intensity to detect EBs. Other studies use clustering techniques to detect EB candidates \citep{2020A&A...641L...5J,2021A&A...648A..54R, 2022A&A...664A..72J, 2024arXiv240609585B}. They used the $k$-means algorithm \citep{1056489} to cluster spectral profiles into groups with similar properties. Once the data is sorted into the different clusters, one has to manually select which clusters contain representative properties of EBs. This method has the advantage of using all the information encoded in the spectral line; however, it does not provide a probability of belonging to a particular group or category.

Although the mentioned methods work well for detecting EBs, they are typically tailored to individual observations and are not easily applicable across different datasets. Additionally, the observations where EBs are detected are usually conducted using ground-based telescopes, which generally cover a small FOV and have limited temporal coverage. This poses a significant limitation for studying EBs over larger spatial and temporal domains, as well as for understanding their impact on the long-term evolution of active regions. One potential solution to this problem is detecting EBs in the passbands of the Solar Dynamics Observatory's Atmospheric Imaging Assembly (SDO/AIA) at 1600\,\AA\ and 1700\,\AA. This approach could open a new avenue for studying the role of small-scale magnetic reconnection events over longer periods and across the full solar disk. However, detecting EBs in SDO/AIA presents significant challenges. Firstly, the pixel resolution of SDO/AIA is 0\farcs6 \citep{2012SoPh..275...17L}, which means that the telescope will only capture the larger EBs. Secondly, SDO/AIA provides passband images instead of detailed spectral information. This is a major drawback because spectral lines have key features for detecting EBs. An additional difficulty is the similarity between EBs and other phenomena such as magnetic concentrations \citep[also referred to as pseudo EBs in][]{2013JPhCS.440a2007R}. \cite{2019A&A...626A...4V} attempted to extend EB detection to the SDO/AIA 1600\,\AA\ and 1700\,\AA\ passbands on 10 different datasets, using an approach primarily based on intensity thresholding. They used EB detections from co-aligned $\Ha$ observations taken by the Swedish 1-m Solar Telescope \citep[SST;][]{2003SPIE.4853..341S} as ground-truth data. However, their result showed that the use of the 1600\,\AA\ or 1700\,\AA\ intensity channels is not enough to precisely recover all the EBs present in the data. This highlights the considerable challenge of this task, suggesting the need for more advanced methods or alternative techniques that combine all the different data available in the observations.

Deep learning techniques have been extensively used to study the Sun, as reviewed by \cite{2023LRSP...20....4A}. Among other achievements \citep[e.g.][]{2022A&A...659A.165D}, these techniques -- based on the optimization of complex neural networks (NNs) -- have recently demonstrated their effectiveness in automatically detecting various features in solar images, such as coronal holes \citep{2018MNRAS.481.5014I,2021A&A...652A..13J}, solar filaments \citep{2019SoPh..294..117Z,2022SoPh..297..104G}, and sub-granular structures \citep{2022FrASS...9.6632D}. Other studies have used them to detect and differentiate between various solar phenomena such as flare ribbons, prominences, and sunspots \citep[e.g.][]{2019SoPh..294...80A}. Motivated by those advances and the flexibility of these techniques to combine diverse wavelength channels in a non-linear way, we developed a method to identify EBs in $\Ha$ observations from the SST and in SDO/AIA datasets using deep learning. We aim to create models that are generalizable, meaning that they are applicable to any other dataset from the same observatory once they are trained. Moreover, by testing the method in different scenarios, we can quantify the importance of the input information for detecting EBs. 

The paper is organized as follows. Section~\ref{sec:methods} explains the methodology used to build the models, covering the approach taken, the data used, and the specific architecture and assessment of the models. Results are presented in Sect.~\ref{sec:results} and discussed in Sect.~\ref{sec:discussion}.

\section{Semantic segmentation of EBs}\label{sec:methods}

\subsection{Segmentation process}

Our objective is to determine whether a pixel belongs to an EB or not. This task of assigning a label to each pixel of the image based on the object it belongs to is called semantic segmentation \citep{prince2023understanding, 2023arXiv230206378C}. To perform the segmentation, we use methods based on NNs \citep{Goodfellow-et-al-2016}. The primary motivation for using NNs in this work is their capability to uncover deeper patterns and relationships within the data, potentially leading to more effective solutions than traditional methods \citep[see e.g.][]{2023A&A...673A..35D,2024arXiv240905156D}. Additionally, NNs allow the incorporation of a wider range of features, including spatial information, spectral information, and the observer's line-of-sight information.

\begin{figure}[t]
\centering
\includegraphics[width=1\linewidth]{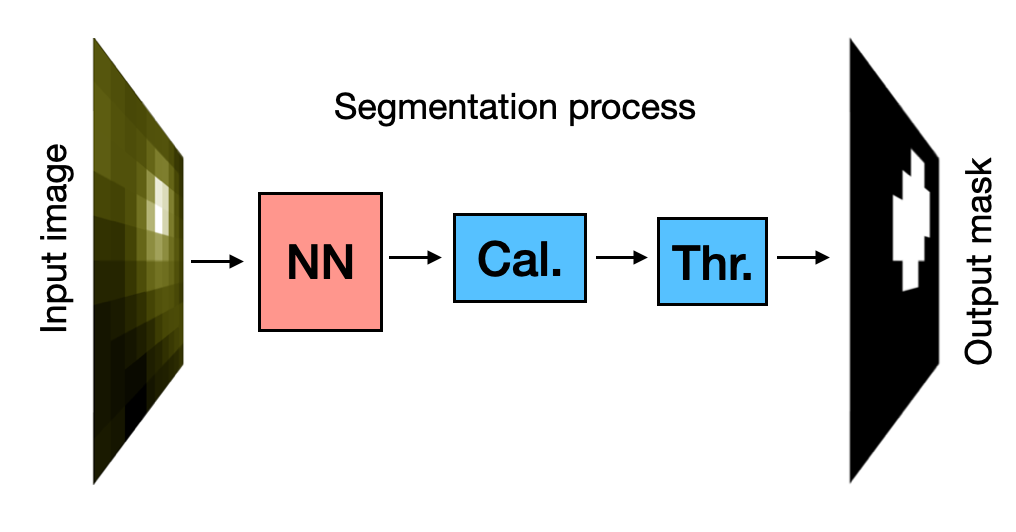}
\caption{Sketch of the semantic segmentation process for the determination whether a pixel in the input image belongs to an EB. After being processed by the neural network (NN), the calibrator (Cal.), and application of a probability threshold (Thr.), the output mask shows which pixels belong to an EB (white).
}
\label{fig:segmentation process}
\end{figure}

A sketch of the full semantic segmentation process is depicted in Fig.~\ref{fig:segmentation process}. From left to right, the input image is processed by the NN, which outputs a value between 0 and 1 for each pixel. This output is refined by a calibrator that adjusts each value to a probability of a given pixel being an EB. Finally, we apply a threshold to these probabilities: all pixels with values above the selected probability threshold are classified as EBs. \textcolor{blue}{From now on, we will refer to this probability threshold simply as probability.}

\subsection{Ground-truth data}

To optimize a NN for any given task, we need ground-truth data, i.e., observations where EBs have been pre-detected and each pixel is already labeled. Part of this data is also used later to assess the model's performance. Our goal is to develop a model capable of generalization, meaning it can be applied to observations from the same telescope but targeting different regions of the Sun. To achieve this, the ground-truth data must be as diverse as possible. Since our objective is to detect EBs in both SST and SDO observations, we need two different groups of datasets. One group with SST observations and the other one with SDO/AIA observations. Each group of datasets is composed of several observations, as explained in Sect.~\ref{subsection:SST_data} and Sect.~\ref{subsection:SDO_data}. We also want these datasets to be co-spatial and co-temporal between SST and SDO/AIA, so we can translate the SST EB $\Ha$ signatures to SDO/AIA channels.

\subsubsection{SST datasets}\label{subsection:SST_data}

The SST dataset is composed of data labeled by \cite{2019A&A...626A...4V}, where EBs were detected in SST $\Ha$ high-spatial resolution observations. The main characteristics of these datasets used are summarized in Table~1 of that paper, and we adopted the same numbering to refer to each different observation. The criteria used in \cite{2019A&A...626A...4V} to classify a candidate as an EB involved a brightness threshold of 145\% for the EB's core and 140$\%$ for its halo over the average quiet-Sun in at least one of the wings of the $\Ha$ line. Regarding the size and lifetime, a minimum area of $\sim 0.035\,\mathrm{arcsec}^2$ and a minimum lifetime of 60\,s were considered. These datasets cover a wide range of heliocentric angles ($\theta$), with $\mu=\cos\theta$ ranging from 0.93 to 0.2. They also include three main types of regions: moat flows, decaying active regions, and emerging flux regions. This variety suits the requirements for diverse data. The spectral coverage of each observation differs, ranging from $\pm 1.4\,\AA$ to $\pm 2.1\,\AA$ with respect to the $\Ha$ line core at $6563\,\AA$.

To combine the data from different observations, we had to move all the data into a common frame. First, we normalized each dataset with the average intensity of all the quiet-Sun pixels for each observation, following \cite{2019A&A...626A...4V}. We opted to keep only the observations with a spectral coverage up to $\pm 1.5\,\AA$ from $\Ha$ line core to avoid possible confusion with network bright points \citep{2013JPhCS.440a2007R}. We then interpolated the data to the following wavelengths with respect to the line core 6563$\,\AA$: [$\pm$1.5, $\pm$1.2, $\pm$1.0, $\pm$0.8, $\pm$0.6, $\pm$0.3, $\pm$0.2,  0.0]$\,\AA$, to ensure the same spectral points across all the observations. This process excluded sets B and F from \cite{2019A&A...626A...4V} because their spectral sampling does not reach $\pm1.5\,\AA$. We also excluded set C because it was not available. Therefore, we used sets A, D, E, G, H, I, and J. The input features used for SST detections include all the spectral points of the $\Ha$ line. We also experimented with the inclusion of spatial information and the line-of-sight parameter $\mu$ to study the quality of the classification. All the SST data that we analyzed were obtained with the CRisp Imaging Spectropolarimeter \citep[CRISP;][] {2008ApJ...689L..69S}, with a pixel resolution of 0\farcs057~pixel$^{-1}$. Those datasets have been processed with different versions of the SST data reduction pipeline \citep{2015A&A...573A..40D} 
including the Multi-Object Multi-Frame Blind Deconvolution image restoration technique \citep[MOMFBD;][]{2005SoPh..228..191V}.
To illustrate how the detections and the analysis would be carried out in a real-world context, we applied our trained models on a new unseen dataset which lacks ground-truth labels. This dataset consists of an $\Ha$ observation of the AR 13679 observed at the SST. This observation was carried out on 2024 May 15 from 08:44:46 to 09:11:57~UT. We used the $\Ha\,6563\,\AA$ scans, obtained with the new CRISP cameras with a pixel scale of 0\farcs044~pixel$^{-1}$. The line was sampled at 33 positions in the range $\pm2\,\AA$ with respect to the line core at $6563\,\AA$, with 0.1$\,\AA$ steps. The cadence of the instrument for this configuration was 37\,s (the scan also included the \ion{Fe}{I} 6173$\,\AA$ and \ion{Ca}{II} 8542$\,\AA$ lines, but we do not use them here). The dataset was processed using the SSTRED reduction pipeline \citep{2021A&A...653A..68L}. We will refer to this dataset as the \textit{new dataset}.

\subsubsection{SDO datasets}\label{subsection:SDO_data}

The channels we used to detect EBs in SDO/AIA are the 1600\,\AA, 1700\,\AA, 171\,\AA, and 304\,\AA\ passbands. The first two passbands present clear EBs signatures. Additionally, the 1600\,\AA\ passband also displays signatures from flaring active region fibrils and transition-region transients due to a large \ion{C}{iv} line contribution \citep{2018SSRv..214..120Y, 2019A&A...626A...4V, 2019ApJ...870..114S}. These signatures are also visible in 171$\,\AA$ and 304$\,\AA$, so we used these passbands to discard false positives in the 1600$\,\AA$ channel. Therefore, we used these four passbands as inputs for the SDO/AIA model. Futhermore, similar to our approach with SST, we also investigated the inclusion of spatial information and the viewing parameter $\mu$.

\begin{figure*}
\centering
\includegraphics[width=0.8\linewidth]{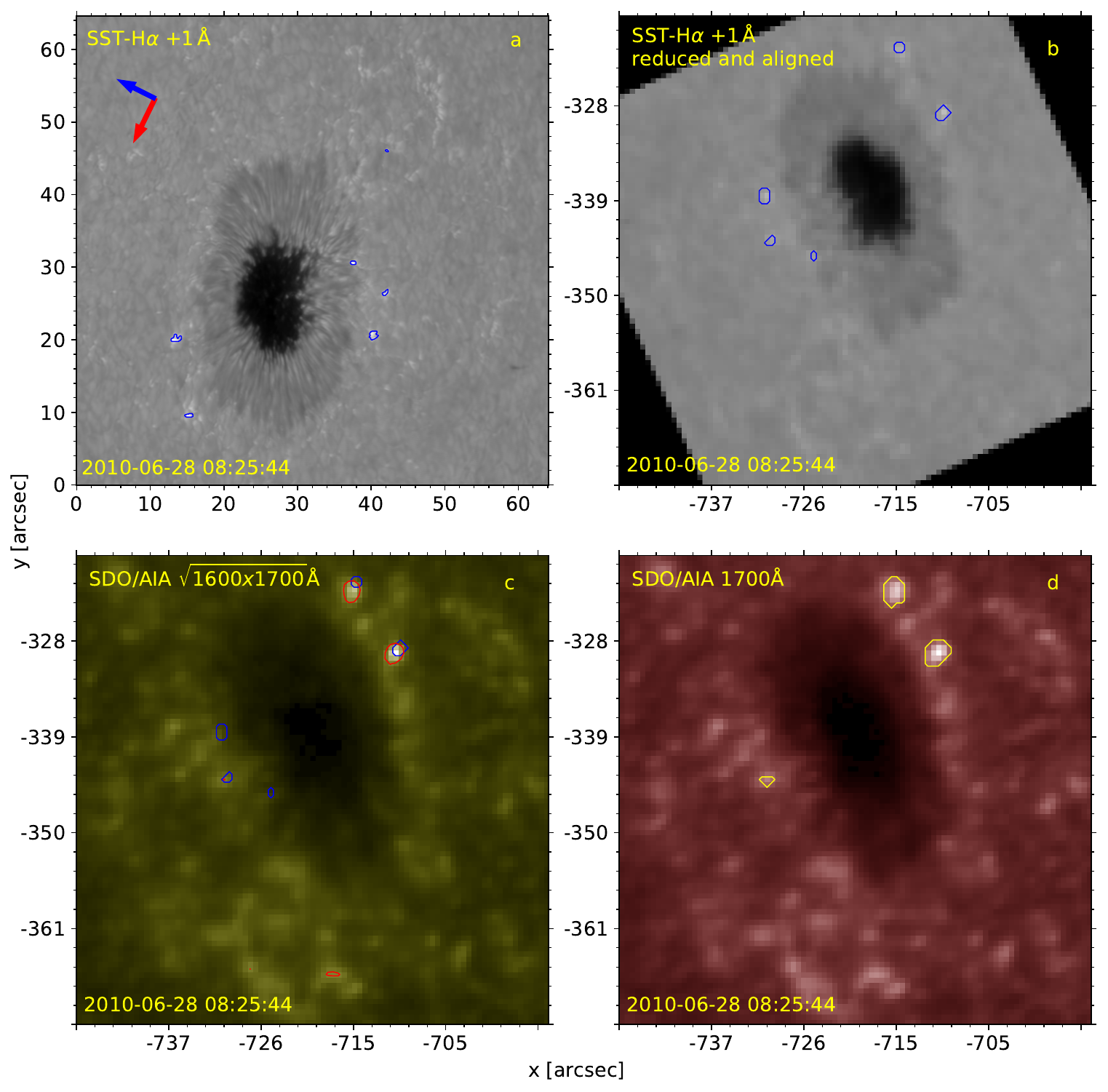}
\caption{Illustration of the process used to compute SDO/AIA labeled data for an arbitrary frame of dataset E.
Panel (a) displays the SST $\Ha$ blue wing and EBs highlighted by blue contours.
Panel (b) shows the same data as the previous panel but downsampled and coaligned with SDO data of the same target region. Panel (c) shows the square root of the product of SDO/AIA $1600\,\AA$ and $1700\,\AA$. Red contours indicate pixels above the threshold and blue contours represent the EBs from the SST. Panel (d) shows SDO/AIA $1700\,\AA$ passband for the target region with the resulting masks in yellow contours. These yellow contours are selected as the red contours in panel (c) which overlap with the blue contours. 
Panel (a) is displayed using its data coordinate system. The arrows at the top left of this panel point towards solar north (red), and west (blue) for this reference frame.
Panels (b), (c), and (d) are displayed using heliocentric coordinates.
}
\label{fig:SST2SDO}
\end{figure*}

Unlike the SST case, we do not have pre-labeled SDO data with detected EBs. Therefore, we have to label SDO data based on SST observations. To achieve this, we downloaded co-temporal SDO/AIA observations for each SST dataset. These observations interpolated to the same pixel scale of 0\farcs6~pixel$^{-1}$, normalized by the exposure time and corrected from time-dependent degradation using the {\tt AIAPY} python package \citep{Barnes2020}. Then, we made cutouts of the corresponding regions to obtain co-spatial SDO/AIA data that matched SST observations. The process to create the masks is illustrated in Fig.~\ref{fig:SST2SDO}. First, we downsampled the SST data and masks by means of pixel binning to match the pixel resolution of SDO/AIA  (from $\sim$0\farcs06 to $\sim$0\farcs6). Next, we coaligned the SST data with the SDO data using the $\Ha$ blue wing and the SDO/AIA 1700\,\AA\ passband (panels a and b), so SST \Ha\ EB masks can be applied to the SDO data. Given the lower resolution of SDO/AIA, only the brighter EBs are visible in the UV passbands \citep[see e.g.][]{2019A&A...626A...4V}. Hence, smaller EBs will not show any signature in UV channels. If we directly use the SST masks as they are, we risk considering SDO pixels as EBs when they do not show any signature, reducing the reliability of the method. Therefore, we had to refine the mask definition for SDO. To re-identify signatures of EBs in UV channels, we search for UV brightenings that overlaped with the EB masks.

To identify these UV brightenings, we applied an intensity threshold to the SDO datasets using the square root of the product of the 1600$\,\AA$ and 1700$\,\AA$ channels. We used this combination to enhance the signal of the brightenings. The value for the intensity threshold is different for each dataset and has been fine-tuned to isolate the brighter regions as indicated by the red contours on panel c of Fig.~\ref{fig:SST2SDO}. Finally, we selected the UV brightenings contours that overlap with EB masks from the SST as final SDO mask, i.e., the red contours that coincide with blue contours on panel c of Fig.~\ref{fig:SST2SDO}. This mask is shown on panel d in yellow.

\begin{figure*}[t]
\centering
\includegraphics[width=\linewidth]{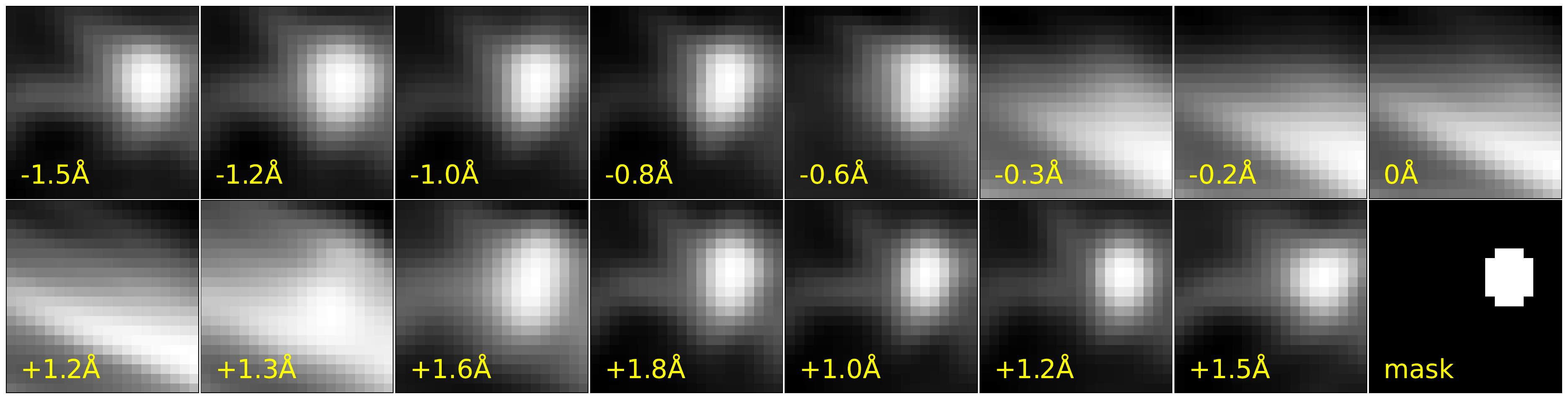}
\caption{Example SST image patch with all the channels. All this information forms one SST image patch, which serves as input for the CNN SST models. Each image is scaled individually. Offsets are given with respect to nominal \Ha\ line center. The size of the area of SST is 20$\times$20 pixels.}
\label{fig:cards_example_SST}
\end{figure*}

\begin{figure*}[t]
\centering
\includegraphics[width=0.8\linewidth]{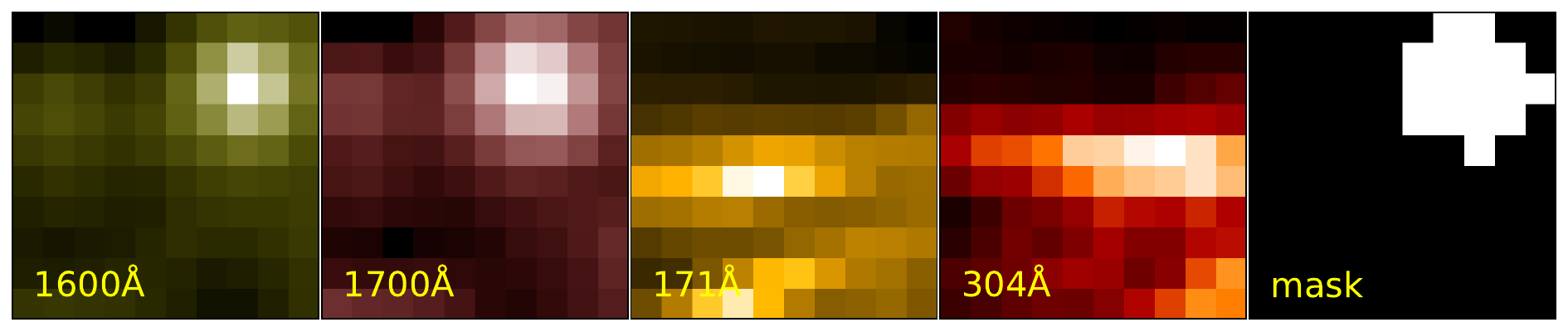}
\caption{Example SDO image patch with all the AIA channels included for the segmentation process, which serves as input for the CNN SDO models. Here the EB is clearly visible in 1600\,\AA\ and 1700\,\AA\ while there are no signatures in the 171\,\AA\ or 304\,\AA\ passbands. The size of the area of SDO image patch is 10$\times$10 pixels.}
\label{fig:cards_example_SDO}
\end{figure*}

\subsection{Training, validation, and test set}
\label{sec:training}

To build and assess the models we used three different datasets: training, validation and test. Training and validation sets were used during the model training process. The training set was employed in the optimization of the NN, while the validation set was utilized to evaluate the model's ability to generalize rather than memorize the labels from the training data. Once the models were built, the test set was used to assess their performance using ground-truth data which was not incorporated during the training process.

To create the training, validation, and test sets, we carefully selected data that include instances of all events seen in the observations (e.g., plages, bright points, flaring (arch)filaments, umbras). This helps the model to distinguish EBs from other phenomena. Without this step, the models may only correctly classify clear-cut cases of EBs or non-EBs and struggle with brighter events like magnetic bright points or flare ribbons.

To create the datasets we used subsamples of the data with dimensions $(m, n, f)$,  where $m$ and $n$ are the spatial dimensions and $f$ includes the spectral information and additional features. We refer to these subsamples as image patches and we used square image patches ($m$ = $n$) in the following. The advantage of using image patches is that they can capture whole events or parts of them, unlike single pixels. Each image patch includes all image channels and spectral positions from their respective observation. This is shown in Fig.~\ref{fig:cards_example_SST} for an arbitrary image patch of the SST dataset and in Fig.~\ref{fig:cards_example_SDO} for an arbitrary image patch of the SDO dataset. In both figures, the last panel shows the mask, marking the pixels that belong to an EB candidate. 

A key challenge in this process is automatically selecting the centers of the image patches to ensure an adequate representation of both EBs and other types of events/regions. As discussed, we aim for the image patches to capture EBs along with other phenomena. However, an important issue to address is the strong data imbalance. The data balance $\mathcal R$ is defined as the ratio between the number of valid and invalid events:
\begin{equation}
    {\mathcal{R}} = \frac{\#\rm events}{\#\rm non\mh events}.
    \label{eq:ratio}
\end{equation}

In our case, an event is a pixel that is part of an EB and a non-event a pixel that is not part of an EB. There are significantly more non-events than events in our dataset so $\mathcal{R}$ is very small. This behavior is very common in classification of energetic phenomena \citep[e.g.,][]{2015ApJ...798..135B}. In particular, $\mathcal{R} = 0.001$ for the SST datasets and $\mathcal{R} = 0.003$ for SDO datasets. Randomly selecting positions from those datasets would make it highly unlikely to sample an EB due to this high imbalance. As a result, the model might focus more on non-events rather than the small fraction of EBs. To mitigate this, we developed an algorithm to iteratively select the centers of the image patches based on Eq.~\ref{eq:ratio}. This algorithm, starting in a random location, selects the center of the next image patch based on a probability distribution function $\mathcal{L}$ that is updated at each step to maintain a balanced $\mathcal{R}$, i.e., to keep $\mathcal{R}$ as close to 1 as possible. Additionally, each time an image patch is created, we apply random transformations (rotation, mirroring, and transposing) to avoid having identical or very similar image patches. The algorithm is explained in more detail in Appendix~\ref{appendix:algorithm}.

For both SST and SDO data, we generate 20 image patches per frame, using 20 frames from each of the seven different datasets.
For SST, the image patch size is $20 \times 20$ pixels (generating a total of $1.1\times10^6$ pixels) and $\mathcal{R}$ is 0.185 using the new algorithm. For the SDO dataset, the image patch size is $10 \times 10$ pixels (a total of $2.8\times10^5$ pixels) and $\mathcal{R}$ is 0.117. For both cases, the algorithm for image patches selection have increased $\mathcal{R}$ by two orders of magnitude. 
These new  $\mathcal{R}$ values show the great improvement in the composition of the datasets, although there still being some unbalance to be addressed. This is solved in Sect.~\ref{sec:training_process}.
Once the image patches for each dataset are created, we combined them into a single full dataset, which we split into train (70$\%$) and validation (30$\%$ of the image patches) sets. 
Additionally, we created the test set by generating 10 image patches per frame from 20 frames for each set. 
To avoid data leakage, we used frames that were not included in the training and validation sets to produce the test set. Table~\ref{table:sets_info} summarizes the number of image patches and $\mathcal{R}$ for all the sets produced.

\begin{table}[]
\caption{Description of the training, validation and test set used for the SST and SDO models.}
\begin{tabular}{cccc}
\hline
\textbf{}    & Train                                                                    & Validation                                                               & Test                                                                     \\ \hline
             & \begin{tabular}[c]{@{}c@{}}\# image patches\\ $\mathcal{R}$\end{tabular} & \begin{tabular}[c]{@{}c@{}}\# image patches\\ $\mathcal{R}$\end{tabular} & \begin{tabular}[c]{@{}c@{}}\# image patches\\ $\mathcal{R}$\end{tabular} \\ \hline
{SST} & \begin{tabular}[c]{@{}c@{}}1960\\ 0.185\end{tabular}                     & \begin{tabular}[c]{@{}c@{}}840\\ 0.190\end{tabular}                      & \begin{tabular}[c]{@{}c@{}}1400\\ 0.180\end{tabular}                     \\ \hline
{SDO} & \begin{tabular}[c]{@{}c@{}}1960\\ 0.120\end{tabular}                     & \begin{tabular}[c]{@{}c@{}}840\\ 0.111\end{tabular}                      & \begin{tabular}[c]{@{}c@{}}1400\\ 0.128\end{tabular}                     \\ \hline
\end{tabular}\label{table:sets_info}
\end{table}

\subsection{NN architectures}

To evaluate when the model requires spatial information to detect EB, we employed different NN architectures. When spatial information, including all channels, is essential, we used Convolutional Neural Networks (CNNs). 
CNNs function by concatenating multiple convolutional layers, where the input of each layer is convolved with a matrix called a kernel, producing an output. The weights of these kernels are updated iteratively during training. 
For this task, we specifically used Fully Convolutional Neural Networks \cite[][]{7478072} which only include convolutional layers (including the activation function).
In contrast, when spatial information is not necessary and only spectral information from individual pixels is required, we employed Fully Connected Neural Networks \citep[FNNs;][]{Goodfellow-et-al-2016}. In this approach, each pixel is treated independently, disregarding spatial coherence.

To train these models, we used image patches as input for the CNNs, preserving their spatial structure. For the FNNs, we flatten the image patches into one-dimensional arrays. 
All models used the Exponential Linear Unit  \citep[ELU;][]{2015arXiv151107289C} as the activation function. We selected ELU and not the popular Rectified Linear Unit \citep[ReLU;][]{10.5555/3104322.3104425} because, after testing both, the former proved to be more suitable for the calibration process.
The Sigmoid function was used as the final activation function to map outputs to the range $]0, 1 [$. 

The particular architectures for SST and SDO differ due to their different input features. 
Different architectures were tested, resulting in the ones showed below. Architectures with more or larger hidden layers did not present a large improvement on the final trained models.
For SST, the FNN architecture is:
\begin{equation}
\begin{aligned}
    \overset{\mathrm{Input}(\rm Spectrum)}{15]}&\rightarrow[15,32]\rightarrow\mathrm{ELU}\rightarrow[32,20] \rightarrow\mathrm{ELU}\\
    &\rightarrow[20,10]\rightarrow\mathrm{ELU}\rightarrow[10,1]\\
    &\rightarrow\mathrm{Sigmoid}\rightarrow\underset{\mathrm{Output}}{[1},
\end{aligned}
\end{equation}
where each tuple $[a, b]$ represents a hidden layer, being $a$ the number of inputs for the layer and $b$ the number of outputs.
For SST, the CNN architecture is:
\begin{equation}
\begin{aligned}
    \overset{\mathrm{Input\,(\textbf{Image patch})}}{20,20,15]}&\rightarrow[15,32,3]\rightarrow\mathrm{ELU}\rightarrow[32,32, 3]\rightarrow\mathrm{ELU}\\
    &\rightarrow[32,12,3]\rightarrow\mathrm{ELU}\rightarrow[12,1,1]\\        &\rightarrow\mathrm{Sigmoid}\rightarrow\underset{\mathrm{Output}}{[20, 20,1}.
\end{aligned}
\end{equation}
Each triplet $[a,\,b,\,c]$ represents the input number of channels $a$, the output number of channels $b$, and the size of the kernel used $c$ for every hidden layer. The input $[a,\,b,\,c]$ is an \textbf{image patch} of size $a\times b$ with $c$ channels. In the cases where we also include $\mu$, the input becomes 16 (15 spectral points + $\mu$). 

For SDO/AIA, the FNN architecture is:
\begin{equation}
\begin{aligned}
    \overset{\mathrm{Input (Spectrum)}}{4]}&\rightarrow[4,10]\rightarrow\mathrm{ELU}\rightarrow[10,10] \rightarrow\mathrm{ELU}\\
    &\rightarrow[10,6]\rightarrow\mathrm{ELU}\rightarrow[6,3]\rightarrow\mathrm{ELU}\rightarrow[3,1]\\
    &\rightarrow\mathrm{Sigmoid}\rightarrow\underset{\mathrm{Output}}{[1}.
\end{aligned}
\end{equation}

For SDO/AIA, the CNN architecture is:
\begin{equation}
\begin{aligned}
    \overset{\mathrm{Input\,(\textbf{Image patch})}}{10,10,4]}&\rightarrow[4,10,3]\rightarrow\mathrm{ELU}\rightarrow[10,50, 3]\rightarrow\mathrm{ELU}\\
    &\rightarrow[50,50,3]\rightarrow\mathrm{ELU}\rightarrow[50,10,3]\rightarrow\mathrm{ELU}\\
    &\rightarrow[10,1,1]\rightarrow\mathrm{Sigmoid}\rightarrow\underset{\mathrm{Output}}{[10,10,1}.
\end{aligned}
\end{equation}
Similar to the case of SST, for SDO, the models where we also included $\mu$, the input and first layers becomes 5 (4 passbands + $\mu$). The models are trained and tested using image patches to reduce computational costs. When making predictions with new observations, it is not necessary to use image patches; the models can be applied directly to the full-resolution data.

We tested more sophisticated architectures, but they did not offer any improvement over the models we ultimately selected. This outcome suggests that incorporating an encoder/decoder architecture, typical in U-Net models \citep{2015arXiv150504597R,2019A&A...629A..99D}, would not offer additional benefits.

\subsection{Training process}\label{sec:training_process}

The models were trained for several hundreds of epochs (around 500) until they no longer improved. The batch size was chosen to be 126 for the SST and 500 for the SDO to achieve a good balance between memory consumption and performance.
We used the Adam optimizer \citep{kingma2017adam} for training the neural networks. Different learning rates were tested for each model, selecting the ones that presented the best performance. All the selected learning rates are between $10^{-2}$ and $10^{-3}$. We used a weighted Binary Cross-Entropy (BCE) as the loss function to address the class imbalance in the data:
\begin{equation}
        \text{BCE} = \sum_{n}^N -w_n\left[ y_n\log{\hat{y}_n} + (1 - y_n)\log(1-\hat{y}_n)\right],
\end{equation}
where $\hat{y}_n$ is the output of the NN model and $y_n$ is the ground-truth data for the $n_{th}$ pixel, and $N$ is the total number of pixels. The BCE is a loss function used in binary classification tasks that measure the difference between the ground-truth labels and the predicted probabilities, employing logarithms to penalize incorrect predictions more severely. Unlike Mean Squared Error (MSE), which is used for regression, BCE is tailored for classification, effectively handling the probabilistic discrepancy between predicted and actual class labels.
The weights ($w_n$) were employed because models trained with the plain Binary Cross-entropy tend to develop a bias toward the majority class in imbalanced datasets. By incorporating these weights, we assigned different levels of importance to event and non-event pixels,  ensuring a balanced training process \citep{Goodfellow-et-al-2016, 2023arXiv230206378C}. 
The weight for each class was computed using the formula:
\begin{equation}\label{eq:weights}
    w_{c} = \frac{1/N_c}{\sum_c^C 1/N_c},
\end{equation}
where $N_c$ is the number of pixels of class $c$, and $C$ denotes the total number of classes, which is 2 in our case. Each pixel is assigned a weight based on its class. During the training, the weight of each pixel within a batch is normalized by the sum of the weights of all pixels in that batch. One might assume that these weights alone would adequately address any class imbalance issue; however, the original imbalance was so extreme that a combination of weighted BCE and the implemented resampling techniques, as described in Sec.~\ref{sec:training}, was necessary to achieve satisfactory performance.

\subsection{Calibration}\label{sec:calibration}

After obtaining the NN output, we proceed with the calibration process (see Fig.~\ref{fig:segmentation process}). The output of the NN is a number between 0 and 1 that we would like to interpret as the probability of being an EB. For instance, an output of 0.6 should reflect a 60$\%$ chance of belonging to the EB class. If the model's output closely matches to the actual probability, the model is considered calibrated. Although the loss function used in the training process includes a probabilistic interpretation, the NN output is often not calibrated \citep{DBLP:journals/corr/GuoPSW17}. 

The calibration process of a NN can be addressed as a post-processing step. The most common method consists of fitting a function that maps the output of the NN to a true probability. Following \cite{DBLP:journals/corr/GuoPSW17}, we chose to fit a isotonic step-wise function \citep{kruskal1964nonmetric} to calibrate the models. This function is a non-parametric regression that maps the output of the NN to the true probability. The calibration process is performed using the training and validation sets together because the calibration process requires as much data as possible to sample all the possible outputs. The calibration process does not affect the performance of the model, but it significantly improves the interpretability of the predictions. 
The calibration process is explained in more detail in Appendix~\ref{calibration_apendix}.

\subsection{Probability}\label{subsec:threshold}

After the calibration process, the NN produces a continuous output in the range $]0,1[$ for each pixel, which can be interpreted as probabilities. To determine the final decision on whether a pixel is an EB or not, a probability value has to be chosen. Pixels with values above this probability will be considered as EB. For instance, by selecting a probability of $0.8$ (indicating an 80$\%$ probability of being an EB), all the pixels with a probability of 0.8 or higher will be classified as EB. This final step, shown in Fig.~\ref{fig:segmentation process} as \textit{Thr.}, results in the output binary mask,  where 1 indicates EB, and 0 indicates non-EB.

\subsection{Performance metrics}

After building and training the model, we evaluated its performance on the test set. This evaluation involved comparing the model's predictions $\hat{y}$ to the ground-truth labels $y$ in the validation set. The predictions were classified into the following categories:
\begin{itemize}
    \item True Positive (TP): the model correctly predicts an event that is indeed present. This means that both the predicted label and the actual label are 1 ($y = \hat{y} = 1$).
    \item False Positive (FP): the model incorrectly predicts an event that is not present. In this case, the actual label is 0, but the predicted label is 1 ($y = 0$ but $\hat{y} = 1$).
    \item True Negative (TN): the model correctly predicts that there is no event. Here, both the predicted label and the actual label are 0 ($y = \hat{y} = 0$).
    \item False Negative (FN): the model fails to predict an event that is actually present. In this scenario, the actual label is 1, but the predicted label is 0 ($y = 1$ but $\hat{y} = 0$).
\end{itemize}
These four quantities are considered as the elements of the confusion matrix or contingency table of any binary classification task. 

By categorizing the predictions in this manner, several metrics can be computed. For example, recall measures how many events are recovered over the total number of true events:
\begin{equation}
    \rm recall = \frac{\rm TP}{TP+FN} .
\end{equation}\label{recall}
A recall of 1 means that we correctly classified all the EBs present in the data.

The precision measures the fraction of correct predictions over the total number of predicted events: 
\begin{equation}
    \rm precision = \frac{TP}{TP + FP}.
\end{equation}
A precision of 1 means that among all the predictions of EBs there is no prediction that is not an actual EB.
Recall and precision are anti-correlated except in the case of a perfect classifier. If the precision of a model is increased, the model becomes more conservative to reduce the chances of misclassifying the positive class, thus decreasing the ability of the model to recover all the positive examples (i.e., the recall), and vice versa. Therefore, a useful quantity to balance recall and precision is their harmonic mean, the $F_1$ score:
\begin{equation}
F_1 = \rm 2\frac{precision \cdot recall}{precision + recall}.
\end{equation}

The $F_1$ score and the precision are metrics that are strongly influenced by the class imbalance in the datasets. Therefore, they should not be used to compare models that are trained on datasets with different $\mathcal{R}$. However, since we are comparing models trained with the same dataset, this does not pose an issue for our comparison. Although more complex metrics exist \citep[see e.g.,][]{2015ApJ...798..135B, 2023LRSP...20....4A}, we decided to use these fundamental metrics due to their simplicity and ease of interpretation.

\subsection{Spatial degradation}

To investigate the impact of a lower spatial resolution on the detection of EB, we degraded the SST data to match the spatial pixel size of SDO/AIA, which is ten times larger.
To achieve this, we first downsampled the SST data to the SDO pixel size and then upsampled it back to the original SST dimensions using nearest-neighbor interpolation, ensuring the size of the original EB masks was preserved.

By creating these degraded datasets, we could systematically assess how the reduced spatial resolution affected the performance of our EB detection models. This step is crucial to disentangle the impact of the spectral and spatial information in this identification. When using SDO, both the spatial resolution and the wavelength information differ from the original SST observations. Therefore, it  is important to quantify how each factor - spatial resolution and spectral characteristics - contributes to the effectiveness of EB detection.

\begin{figure*}
\centering
\includegraphics[width=1\linewidth]{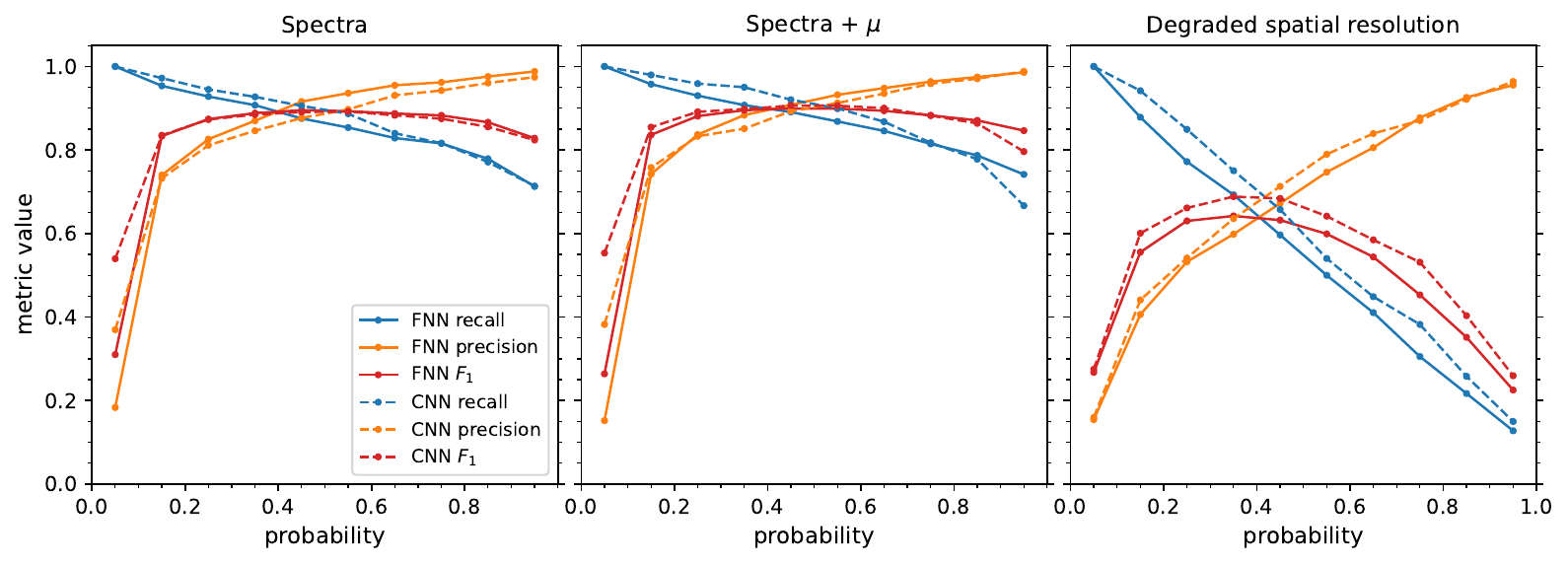}
\caption{Metrics of recall, precision and their harmonic mean, the $F_1$ score, for the models trained to perform EB detection in SST $\Ha$ observations. The metric scores are shown as function of probability of a detection being an EB. From left to right, the metrics are for models trained using: all the spectral points, all spectral points plus $\mu$, and all spectral points for the spatially-degraded dataset. Dashed lines represent CNN models, while solid lines indicate FNN models. Blue, orange, and red colors correspond to recall, precision, and $F_1$ score, respectively. These metrics have been computed using the test set.}
\label{fig:full_metrics_SST}
\end{figure*}

\subsection{Feature importance}

Finally, we conducted an analysis to identify which channels or input characteristics are the most relevant for detecting EBs. For that we used a model inspection technique called permutation feature importance \citep{2018arXiv180101489F}. Permutation feature importance measures the decrease in a model's performance when the values of a single feature are randomly shuffled. This procedure disrupts the relationship between the feature and the target variable, and the resulting decrease in the model's performance indicates how much the model relies on that particular feature.

We used the validation set to calculate the feature importance for the SST and SDO segmentation. For each channel under assessment, we shuffled the values of that specific feature across all pixels while keeping other features unchanged. The model then made predictions using these modified validation sets, and we computed the $F_1$ score to quantify the impact of shuffling each feature.

\section{Results}\label{sec:results}

\subsection{SST EB segmentation}

Figure~\ref{fig:full_metrics_SST} shows the recall, precision, and $F_1$ score and their dependence on the selected probability for all the models trained to detect EB in SST observations. The recall, precision, and $F_1$ score are all functions of the probability threshold applied to the model's output. As the probability increases, the model becomes more restrictive and selects only the most probable events, leaving many real events out of the selection. This is reflected in the recall, which decreases (ratio of recovered events over total events) and the precision, which increases (ratio of correct predictions over total predictions). The $F_1$ score, being the harmonic mean of recall and precision, provides a balanced metric that considers both false positives and false negatives.

Regarding the general trends of different models, the metrics for the CNN and the FNN are very similar in the two first panels of Fig.~\ref{fig:full_metrics_SST}. The recall (blue lines) of the CNN is higher than that of FNN at most points, indicating that the CNN slightly recovers a few more events. In contrast, the FNN models show a higher precision, meaning that they fail less than the CNN when they classifiy a pixel as EB. Despite these differences, the $F_1$ score is almost identical across all probabilities, suggesting similar overall performance for the CNN and FNN models. The inclusion of $\mu$ (second panel) does not seem to add significant information, as the metrics remain similar with or without it.
The maximum $F_1$ score reached in both panels is 0.9, achieved at $40-50\%$ probability, indicating the optimal balance between precision and recall. However, with a $90\%$ probability, the models reach a precision of $\sim0.99$, at the cost of lowering the recall to values between 0.66 to 0.74 depending on the particular model. Thus, although almost all the EB detected are correct, only about a $70\%$ of the present EB in the data are retrieved.
In the third panel, showing the models trained with spatially degraded data, there is a significant difference between CNN and FNN models. The CNN models (dashed lines) achieve better results than FNN models (solid lines) for all three metrics. This behavior indicates that the inclusion of spatial information in the CNN model, clearly improves the detection of EBs. Nonetheless, the maximum $F_1$ score is 0.69 for the CNN and 0.63 for the FNN, at 0.3--0.4 probabilities, 0.27 points lower than the maximum $F_1$ achieved by the models with the original SST resolution. At a 0.9 probability, precisions are 0.95 and 0.96, with recalls of 0.14 and 0.12 for CNN and FNN, respectively.

\begin{figure}
\centering
\includegraphics[width=1\linewidth]{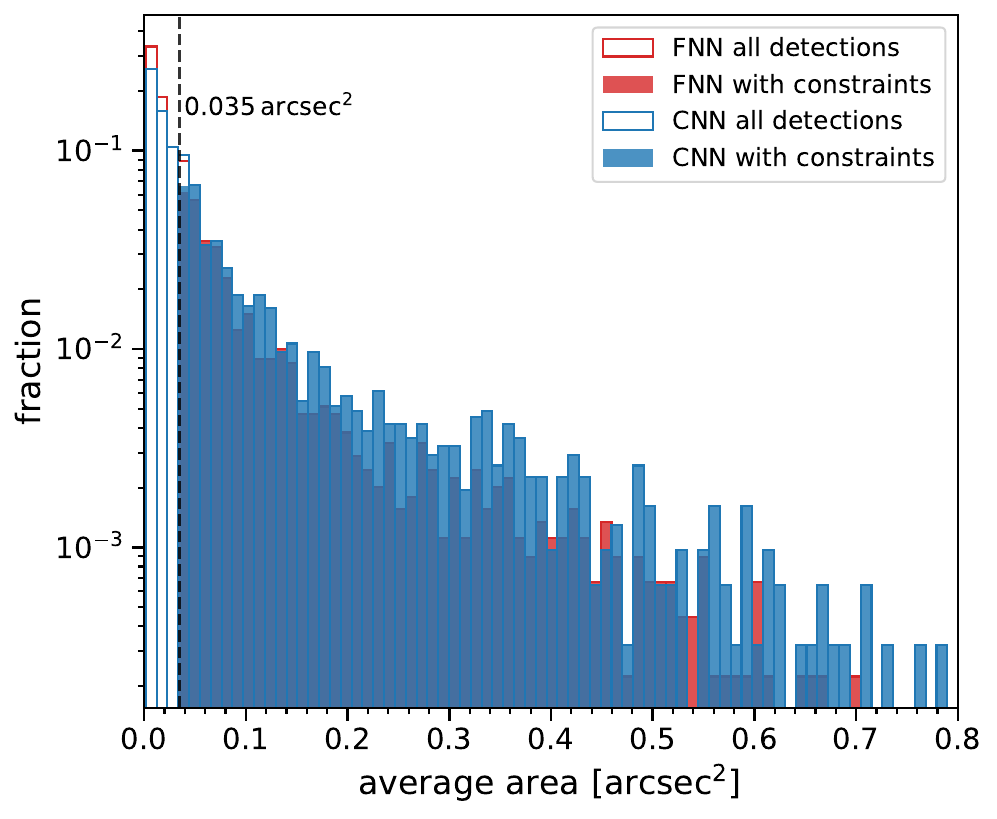}
\caption{
Histograms of the area of EB predictions in the new dataset by the CNN and FNN models, using a probability of 0.5.
Red bars correspond to the detections made by the FNN model and blue bars to the CNN model. Filled bars represent the area of EBs that meet the size criteria explained in the text.
Dark blue color represents the overlap between red and blue bars.
Unfilled bars show all the detected EBs, without size constraints. The dashed line marks the area cutoff used in the criteria. 
The y-axis presents the fraction of each bin with respect to the total number of detections. 
For clarity, the x-axis is limited to 0.80, although the histogram tails extend up to approximately 1.2\,arcsec$^2$ with near-zero counts.
}
\label{fig:histogram_detections_SST}
\end{figure}

To demonstrate the models' ability to handle unseen data, we used them to detect EB in the new dataset.  
Figure~\ref{fig:histogram_detections_SST} shows normalized histograms of the area of the EBs classified in each frame by the CNN and the FNN models, using a probability of 0.5. This value  provides the best trade-off between precision and recall for all three panels in Fig.~\ref{fig:full_metrics_SST}. 
To compute the histograms, we selected detections with an area larger than 0.035~arcsec$^2$ and a minimum linear extent of about 2\arcsec, consistent with the criteria applied in \cite{2019A&A...626A...4V}.
It is evident our models detect smaller features than those identified in \cite{2019A&A...626A...4V}.
The total number of detections are 4483 and 3094 for the FNN and CNN models respectively (without constraints).

The average area for EBs detected with the criteria is 0.118~$\mathrm{arcsec^2}$ for the FNN model and 0.143~$\mathrm{arcsec^2}$ for the CNN model. For all the detected EBs (without the criteria) the average area is 0.050$\,\mathrm{arcsec}^2$ for the FNN model and 0.030$\,\mathrm{arcsec}^2$ for the CNN model. 
It is important to note that we did not apply any temporal criteria to the EB detection. We considered the areas of EBs detected at each frame independently of the other ones. Since EBs increase and decrease in size during their lifetime, all the previous and later smaller sizes after a given EB reaches a maximum size,  displaces the average towards lower values.
The inclusion of temporal evolution could modify the result presented in Fig.~\ref{fig:histogram_detections_SST}, and would reduce the number of detected EBs by connecting them between frames and treating them as single events.

\begin{figure*}[t]
    \centering
    \includegraphics[width=1\linewidth]{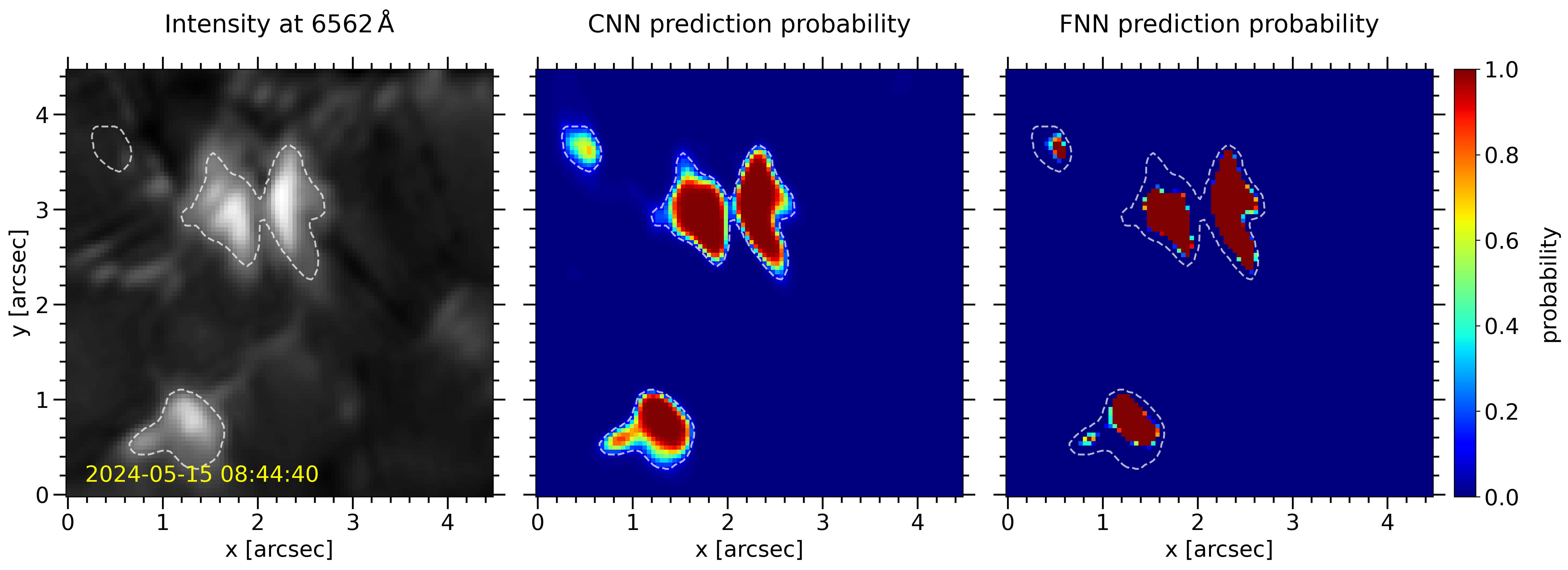}
    \caption{Example of EB predictions done by the CNN and FNN SST models over a cutout of the new dataset. The left panel shows the intensity image in the red wing of $\Ha$ at +1\,\AA\ offset. The middle and right panels show the predictions done by the CNN and FNN models respectively. The colormap indicates the probability assigned by the models to each pixel of being an EB. 
    The white dashed contours mark the 0.1 probability of the CNN prediction and are drawn in all panels to facilitate spatial comparison.
    An animation of this figure that shows the probability maps over the full time sequence is available in the online material.}
    \label{fig:SST_probabilities}
\end{figure*}

Figure~\ref{fig:SST_probabilities} shows a cutout of an SST $\Ha$ frame of the new dataset. The left panel shows the intensity image at +1\,\AA\ from the $\Ha$ core where the EBs appear as intensity enhancements brighter than their surroundings. The middle and right panels show the EBs predictions made by the CNN and the FNN models, respectively, indicating with different colors the probability assigned to each pixel.
CNN detections are smoother than FNN ones. The change from 0$\%$ (dark blue) to 100$\%$ (dark red) is gradual for the CNN while the FNN prediction map shows abrupt changes from 0 to 1.

\begin{figure}
\centering
\includegraphics[width=0.85\linewidth]{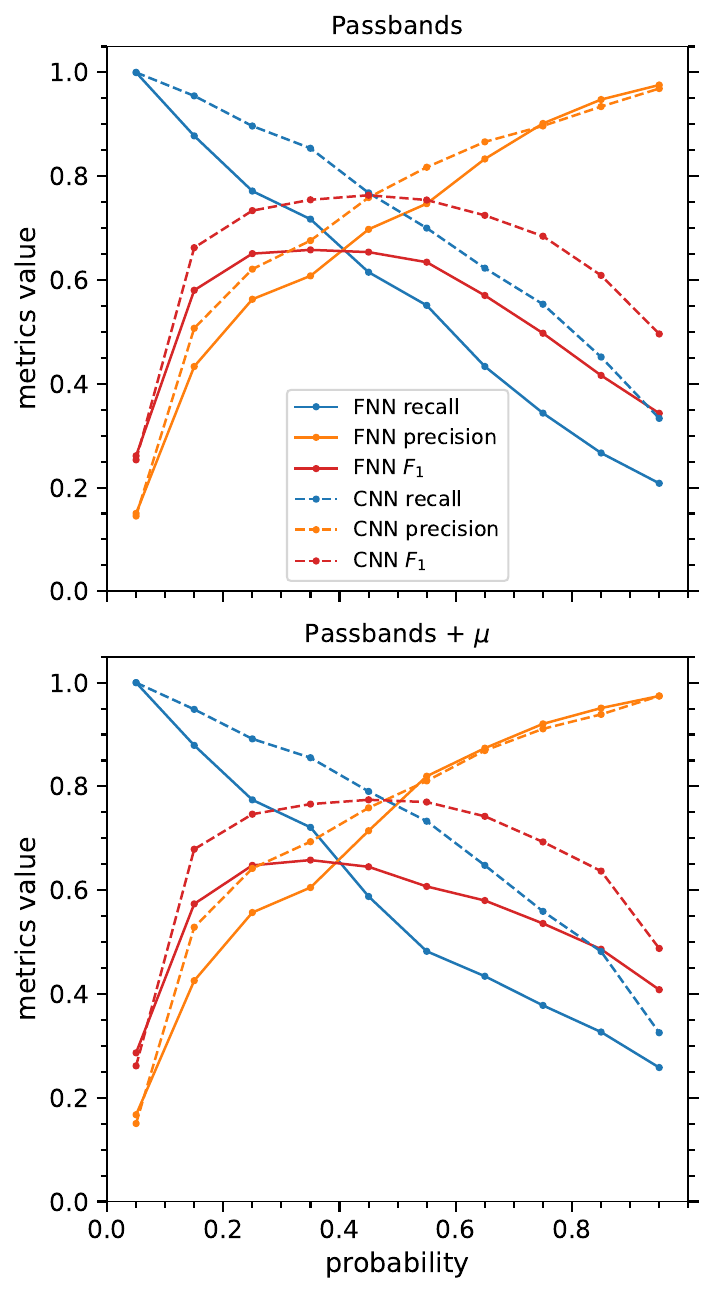}
\caption{
Metrics of recall, precision and $F_1$ score for the models trained to perform EB detection in SDO/AIA. The upper and lower panel shows the metrics for the models trained with all the passbands and passbands plus $\mu$ respectively. The metric scores are shown as function of the probability of a detection being an EB. These metrics have been computed using the test set.}
\label{fig:full_metrics_SDO}
\end{figure}

\subsection{SDO EB segmentation}

Figure~\ref{fig:full_metrics_SDO} presents the performance metrics for SDO/AIA models. The top panel shows the metrics for the models trained using all the passbands, while the bottom panel shows the metrics from the models trained with all the passbands including $\mu$. The inclusion of $\mu$ does not significantly affect the metrics, consistent with the SST results in Fig.~\ref{fig:full_metrics_SST}.

\begin{figure}[t]
\centering
\includegraphics[width=\linewidth]{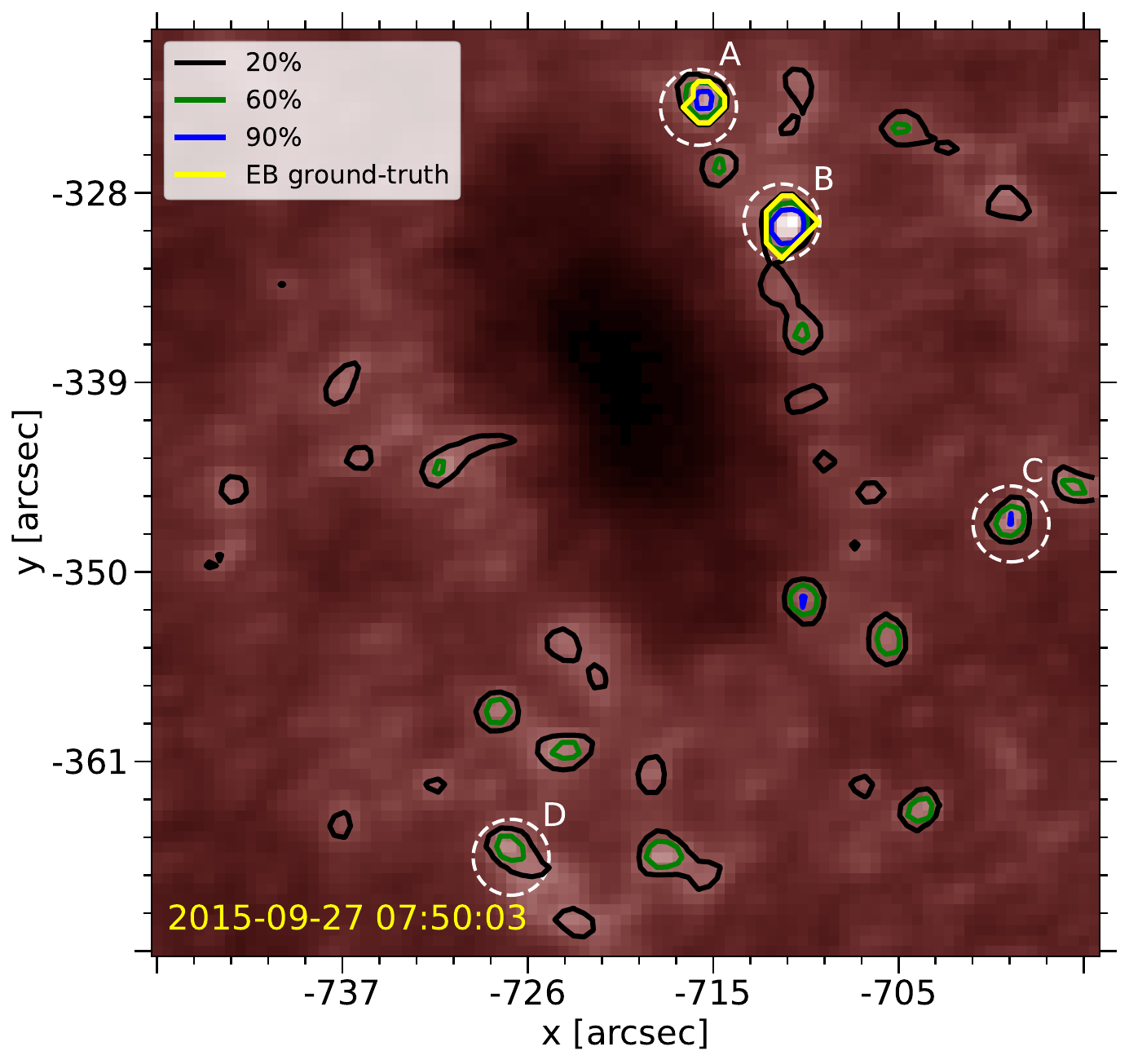}
\caption{
Example AIA 1700\,\AA\ frame with EBs detected by the SDO/AIA CNN model. Detections are outlined by black, green, and blue contours, corresponding to probabilities of 0.2, 0.6, and 0.9 respectively. Yellow contours indicate EBs detected in $\Ha$ SST observations at the same time. White dashed circles and white letters indicate the bright patches selected for the study presented in Fig.~\ref{fig:SDO_lightcurve}. 
The movie online shows the detections for all the frames of the observation.}
\label{fig:SDO_thresholds}
\end{figure}

Comparing the performance of the FNN and CNN models, the CNN metrics (dashed lines) generally achieve better scores across most probabilities. The CNN models achieve a maximum $F_1$ score of approximately 0.77 for a probability of 0.5. The FNN models peak at around 0.66 for a probability of 0.25.
At the highest probability, the precision for both models in both panels reaches around 0.97, with recalls between 0.20 to 0.33. This behavior indicates that the model only recovers between 20--30$\%$ of all the events, missing more than half of them. 
These scores indicate that SDO models cannot accurately recover EBs without making a strong compromise in the number of events detected. 
Finally, Fig.~\ref{fig:SDO_thresholds} shows the predictions made by the CNN model over dataset E. Different contours represent different levels of probability. This illustrates how the model's detection performance varies with different probabilities.

\subsection{Feature importance analysis}

\begin{figure}[t]
\centering
\includegraphics[width=\linewidth]{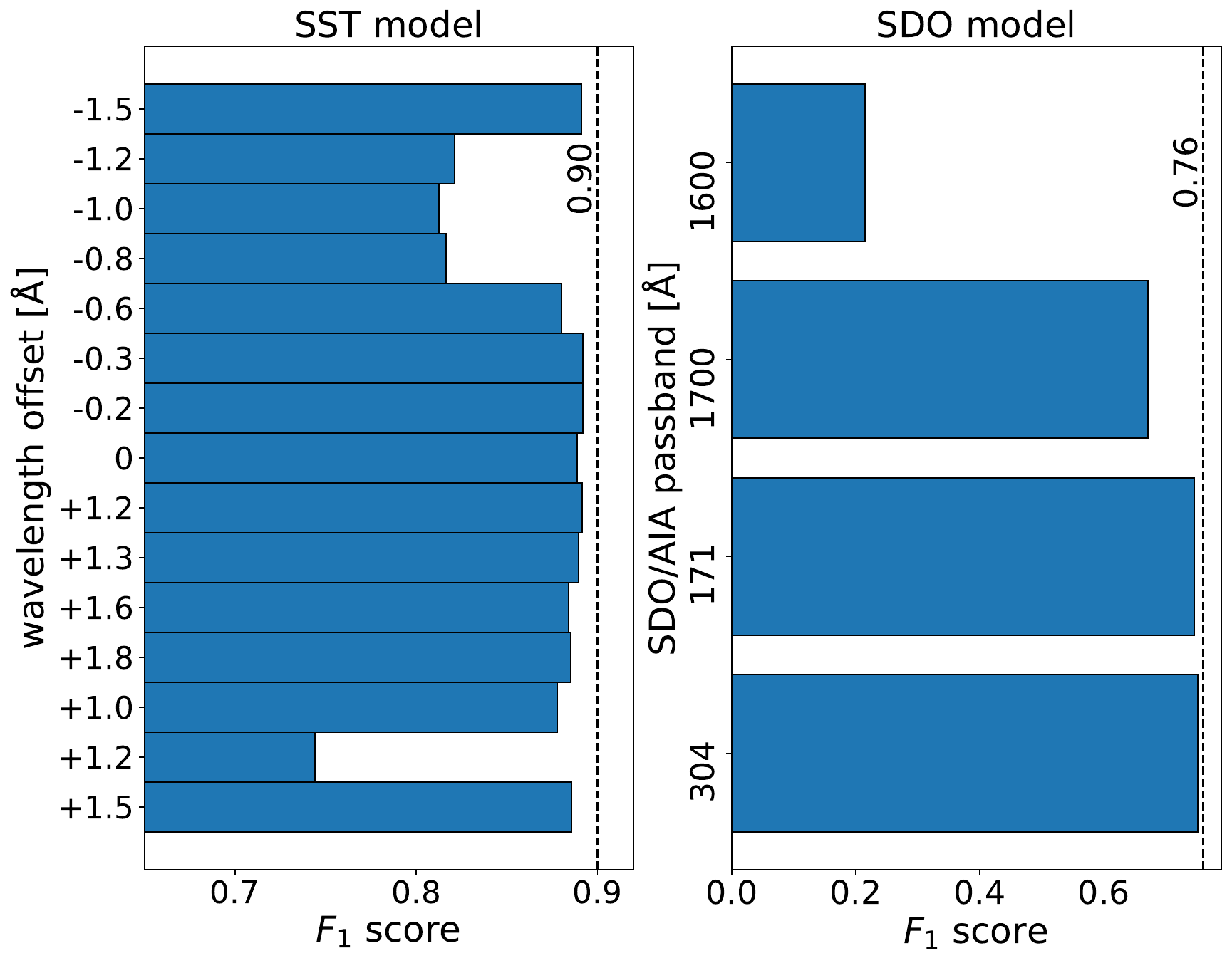}
\caption{$F_1$ score computed for each input channel in the permutation feature importance assessment. The evaluated channel is shuffled to measure performance loss, where a lower score indicates higher importance. The left panel shows the results for the SST input features (wavelength offsets in $\Ha$) and the right panel displays the results for the SDO/AIA passbands. Both panels show the results for the CNN architecture. A dashed vertical line indicates the score without shuffling the input features.}
\label{fig:single channel metric}
\end{figure}

The results of the permutation feature importance experiment are presented in Fig.~\ref{fig:single channel metric}.
This figure shows the impact on the $F_1$ score when the data for each individual wavelength feature is shuffled. As a result, the features for which the $F_1$ score decreases the most are the most important ones. 
We used a 0.5 probability for both SST and SDO CNN models to carry out this section, as this value yields the highest F$_1$ score for the models.
Left and right panels show the $F_1$ score obtained for SST and SDO CNN models, respectively. The dashed lines in both panels indicate the original $F_1$ score without shuffling.
For the SST, data show a sombrero profile, where the scores decrease for the line wings around $\pm 1\,\AA$ offset and increase for the most external points and the line core wavelengths. The feature at $+ 1.2\,\AA$ offset is the most important with a $F_1$ score of 0.74, followed by the blue wing wavelengths at $-1,\,-1.2$ and $-0.8\,\AA$ offsets. The rest of the values, mostly belonging to the line core, lie around 0.89, very close to the score without shuffling (0.9).
We performed the same experiment with the SST FNN model, and the result showed a similar trend, giving a larger importance to the red and blue wings.

In the right panel of Fig.~\ref{fig:single channel metric}, the results for the SDO model show a different relation between the channels.
The lowest score is achieved by the the 1600\,\AA\ channel, with a value of 0.21. The original $F_1$ CNN score without shuffling is 0.76, meaning that the permutation of the 1600$\,\AA$ reduces it by 0.54 points, more than half of the total value. 
The values for 1700\,\AA, 171\,\AA, and 304\,\AA\ are 0.67, 0.74 and 0.75, respectively. 
The gap between 1600\,\AA\ and the rest indicates that 1600\,\AA\ channel is very dominant in the classification compared to the other passbands.

\subsection{EB segmentation vs. intensity threshold}
{To evaluate the effectiveness of the method developed, we aim to compare our trained models with a simple intensity threshold method evaluated in the test set. For the SST models, this comparison is not possible because the ground-truth data used for the training was produced using an intensity threshold in \cite{2019A&A...626A...4V}. Thus, it is redundant by definition to compare the SST models with the intensity threshold method using the test set. For the SDO models, this comparison is possible because we have re-defined our own ground-truth data as explained in Sect.~\ref{subsection:SDO_data}.}

\begin{figure}
\centering
\includegraphics[width=0.9\linewidth]{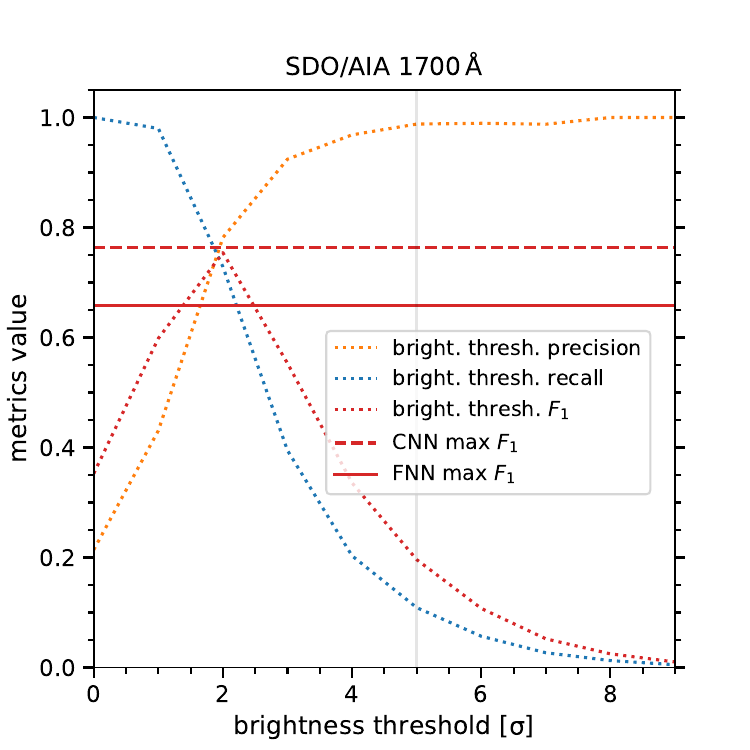}
\caption{Recall, precision and $F_1$ score for the brightness intensity threshold used to perform EB detection in SDO/AIA.
The metric scores are shown as function of the applied intensity threshold. The units of the thresholds is the number of standard deviation values ($\sigma$) above the local quiet-Sun average. For comparison, dashed and solid red lines indicate the maximum $F_1$ score achieved by the FNN and CNN SDO models respectively. The gray vertical line indicates the threshold value suggested by \cite{2019A&A...626A...4V} at 5$\sigma$. These metrics have been computed using the test set.}
\label{fig:brightness_threshold}
\end{figure}

To evaluate the effectiveness of our SDO models, we have compared these results with a simple intensity threshold method evaluated in the test set.
We applied a minimum brightness threshold over the 1700$\,\AA$ channel. We used different threshold values, ranging from 1 to 9 standard deviations ($\sigma$) above the local quiet-Sun average, as carried out by \cite{2019A&A...626A...4V}. The result is shown in Fig.~\ref{fig:brightness_threshold}. The best $F_1$ score is 0.75 for a threshold of 2 $\sigma$ over the quiet-Sun average intensity, with a recall and a precision of 0.73 and 0.78 respectively. For comparison, the FNN and the CNN achieved a maximum $F_1$ 0.66 and 0.76. This means that the threshold performs better than the FNN model and is very close to the CNN model.

\section{Discussion}\label{sec:discussion}

\subsection{SST detections}

For the models trained with the datasets at the original SST spatial resolution, both CNN and FNN models show proficient detection and classification of EBs, with very similar performance metrics (see Fig.~\ref{fig:full_metrics_SST}). Notably, the $F_1$ scores for both models are almost identical, with the FNN demonstrating slightly higher precision and lower recall, possibly related to the way both models handle the input information and other differences in the neural network architectures. 
The $F_1$ score value does not vary much for the different probabilities except for 0.1 in the two first panels of Fig.~\ref{fig:full_metrics_SST}.
This indicates that the recall decreases at the same rhythm as the precision increases. As we increase the probability, the models are more restrictive at selecting events to become more precise.
The similarity in the scores between CNN and FNN models indicates that the spectral information is enough to classify EBs, regardless of the spatial information.
However, for the spatially degraded SST observations, CNN models achieve better scores than FNN models, highlighting the importance of having access to the spatial information in this context. During the degradation process, ten pixels are interpolated into one, significantly weakening the EB spectral signature and typical spectral shape. According to \citet{2019A&A...626A...4V}, the typical size of an EB in these datasets is about 0.15~arcsec$^2$. The pixel scale of SST/CRISP is 0\farcs057 so a typical EB will extend around 6.8 pixels. In addition, EBs tend to be elongated, so their signature is greatly affected by the spatial degradation.
The spatial extension of an EB is used by the CNN models to improve identification. As the quality of spectral information decreases, the use of intensity spatial patterns increases.
Regarding the dependency on $\mu$, no substantial differences are observed between models using or not using this information, as shown in the first and second panels of Fig.~\ref{fig:full_metrics_SST}.
These models use spectral points as the primary information, with the spatial information (in the case of CNN) playing a secondary role. The observer's parameter $\mu$ could be relevant if we only use single-channel intensity images, as EB appearance depends on the location on the disk. EBs may appear more distinct towards the limb due to higher contrast and increased projected area, displaying a flame morphology, as explained by \cite{2011ApJ...736...71W} and \cite{2019A&A...626A...4V}.

The inclusion of spatial information in the detection also impacts the properties and quantity of detected EBs in the new dataset. Indeed, Fig.~\ref{fig:histogram_detections_SST} reveals a difference in behavior between CNN and FNN models.
The FNN and CNN models detected a total of 4483 and 3094 EBs respectively, resulting in a difference of 1389 events. The FNN model detected more small-area events than the CNN model, as captured by the size of the bars for the CNN and FNN from around 0.1$\,\mathrm{arcsec}^2$ onwards. In contrast, the FNN model detects more small events in relation to all its detections. 
The CNN model imposes some spatial coherence and is more restrictive for small candidates,  indicating that the CNN model also incorporates some spatial information (from the training set) when detecting EBs. 
The EBs present in the ground-truth data have a minimum area constraint of 0.035\,arcsec$^2$, which is softly learned by the CNN model during the training process. Since the spectral information dominates, the size requirement is probably applied by the model when the predictions are dubious. 
However, the differences between the CNN and FNN detections are not very significant. For the FNN, 89\% of all detections are below 0.1$\,\mathrm{arcsec}^2$, while for the CNN, they are about 80\% of the total. These conclusions are specific to this dataset and may not be generalizable.

To compare the areas of the EBs detected by our models with the ones reported in \cite{2019A&A...626A...4V}, we applied the same area and linear extent constraints, represented by the color-filled bars in Fig.~\ref{fig:histogram_detections_SST}. With this, we got an average EB area of 0.14 and 0.12$\,\mathrm{arcsec}^2$ for the CNN and FNN models. This agrees with the average EB area of 0.14\,arcsec$^2$ found by \cite{2019A&A...626A...4V}. However, if we remove the size constraints for the detected EBs, we obtain average areas of 0.05 and 0.072\,arcsec$^2$ for FNN and CNN models, respectively. Sub 0\farcs1 events also present EB signature. Therefore, depending on the definition, those events can be classified or not as EB. 
As we obtain higher spatial resolution observations, we can expect to find even smaller events showing EB signatures, emphasizing the importance of high-resolution observations.

Figure~\ref{fig:SST_probabilities} shows the probability map predicted by the CNN and the FNN over a cutout of the new dataset. CNN predictions exhibit a smooth transition from non-EB to EB. This smooth transition is very natural since the borders of the EBs are diffuse. In contrast, FNN predictions show abrupt borders, with no probability gradient between 0$\%$ and 100$\%$ except for some single pixels.
These different behaviors are explained by the different architectures. The CNN model's ability to use spatial information allows it to exploit information from adjacent pixels, resulting in smoother transitions in its predictions. On the other hand, the FNN model's pixel-wise approach requires an effective threshold, because it cannot use any extra information. This idea supports why the FNN models classify more small-scale events than the CNN models, as seen in Fig.~\ref{fig:histogram_detections_SST}.


The result of the permutation feature importance experiment for the SST CNN model indicates that the red and blue wings are the most important spectral points to detect EBs.
While the model relies on the entire spectrum for its predictions, the information at the wings of the $\Ha$ line is essential for the correct classification. This result confirms the choice of previous methods based on a combination of both wings for EB detection \citep[see e.g.][]{2013ApJ...774...32V,2015ApJ...812...11V, 2016ApJ...823..110R,2019A&A...626A...4V} and aligns with the spectral EB definition \citep[e.g.][]{2011ApJ...736...71W}. 
The enhancement of the $\Ha$ wings is the main signature of EBs, so it is expected that the wavelengths close to the wings are the most important features for the classification.
However, this does not mean that the other features do not contribute. If the pseudo-continuum wavelengths (farthest wavelength points) or the core are extremely bright, this will tell the model that this is not an EB event. The fact that many of these points show a very small decrease in $F_1$ score when randomized, suggests that we could potentially remove some of them without significantly affecting the model's performance.

\subsection{SDO/AIA detections}

The results for the SDO models are shown in Fig.~\ref{fig:full_metrics_SDO}. The higher $F_1$ values are 0.66 and 0.77 for the FNN and CNN models, respectively. These values indicate that the best performance of these models is poorer than the SST models, although a direct comparison is not  feasible since they are not trained on the same data. 
The $F_1$ score is sensitive to $\mathcal{R}$, so a different $\mathcal{R}$ on the used datasets would scale the obtained metrics \citep{10.1007/978-3-030-50423-6_6}.

Another interesting aspect is that CNN models generally outperform FNN models, and the gap is very well visible with a difference of about 0.1-0.2 in all metrics on average, except for the precision at higher probabilities. This behavior is similar but more pronounced than the one observed for the SST models under the spatially degraded dataset (right panel of Fig.~\ref{fig:full_metrics_SST}). We speculate that this is due to the difference in the spectral information (passbands instead of detailed spectra) and the lower spatial resolution of SDO/AIA (about ten times lower than SST). Our interpretation is that, similar to the SST models under lower spatial resolution, the CNN models use all the available spatial information to detect EBs. The lower spatial resolution of SDO/AIA results in only the largest EBs being visible. Additionally, the lack of detailed spectral information in the SDO/AIA passbands makes it more challenging to differentiate EBs from other bright events.
An example of the predictions made by the CNN model over the SDO/AIA dataset is shown in Fig.~\ref{fig:SDO_thresholds}. At 60\% probability, the CNN model detects a large number of bright patches, which are not EBs. This result indicates that the model struggles to differentiate EBs from other bright events. This behavior is observed in all the frames of this observation (see movie in Fig.~\ref{fig:SDO_thresholds}).

Additionally, the feature importance analysis in Fig.~\ref{fig:single channel metric} shows that the 1600$\,\AA$ passband is the most relevant feature to detect EBs. 
This is consistent with previous findings that the SDO/AIA 1600$\,\AA$ passband shows the highest contrast for EB \citep[e.g.][]{2013JPhCS.440a2007R, 2013ApJ...774...32V}. 
However, the 1700$\,\AA$ passband presents a similar importance as 171$\,\AA$ and 304$\,\AA$ in Fig.~\ref{fig:single channel metric}. 
This conclusion differs from \cite{2019A&A...626A...4V}, which reported that 1700$\,\AA$ passband is better than the 1600$\,\AA$ passband for detecting EBs.  
The difference arises from the detection method used.
In \cite{2019A&A...626A...4V}, single passbands observations were used to detect EBs by applying a threshold on the intensity of each band. Given that
the 1600$\,\AA$ passband presents a contribution of hotter-atmospheric signatures in addition to EBs, the 1700$\,\AA$ passband, which does not have that issue, contains fewer sources of error. This makes the use of the 1700$\,\AA$ passband more precise than the 1600$\,\AA$ passband, as shown in figures 8 and 10 of \cite{2019A&A...626A...4V}.
In the method developed in this work we combined 1600$\,\AA$, 1700$\,\AA$, 171$\,\AA$ and 304$\,\AA$ passbands.
This allows our models to filter out transition-region or flaring events thanks to the information provided by the 171$\,\AA$ and 304$\,\AA$ passbands. 
We speculate that this makes the model to rely on high-contrast brightenings produced by the EBs in the 1600$\,\AA$ passband, making the 1700$\,\AA$ information redundant.
This also indicates that the SDO models apply an intensity threshold over a non-linear combination of the passbands for the classification task. Although the model recovers some EB candidates, this approach is insufficient to accurately discern EBs while keeping a good ratio between the number of events detected and the precision of the detections.

The misidentification of EBs in SDO is mainly due to the similar intensity patterns produced by network bright points and EBs in the SDO/AIA passbands. Network bright points mark strong  magnetic field concentrations, which also appear bright in the wings of $\Ha$ line.
These magnetic field concentrations manifest in the mid-UV channels of SDO as bright patches, which the SDO models also classify as EBs (see Fig.~\ref{fig:SDO_thresholds}). 
The relationship between these magnetic field concentrations and bright patches has been statistically studied by \cite{2022A&A...664A...2T}. 
Most of the $\Ha$ EBs coincide with bright patches in 1600\,\AA\ and 1700\,\AA, that are not clearly distinguishable from those produced by network bright points. Only the strongest EBs induce a noticeable enhancement in the passbands. 
\cite{2013JPhCS.440a2007R} also warned about the misidentification caused by the magnetic field concentrations, referring to them as pseudo-EBs.
Another possible error source is the presence of UV-bursts, which can be mistaken with EBs. As shown in \cite{kleint2022occurrence}, UV-bursts can produce a signature in both AIA 1600\,\AA\ and AIA 1700\,\AA\ passbands, with the 1600\,\AA\ channel statistically displaying a higher enhancement than in 1700$\,\AA$. 
Therefore, using intensity alone is insufficient to detect EBs using SDO/AIA passbands. This work demonstrates that NN-based models are unable to find reliable patterns (spatial or spectral) to differentiate between EBs and other bright events in SDO/AIA data based solely on the intensity maps.

\subsection{Intensity threshold vs. NN models}
{
For the SST case, the use of ground-truth data produced by an intensity threshold does not allow us to compare the NN methods with an intensity threshold. To make the comparison, we would require a new labeled dataset generated independently of the existing ground-truth criteria. However, this would require a new labeling method (with its own bias), making comparisons against a third criterion.}
For the SDO models, we can make the comparison with the intensity threshold method because we created our own ground-truth data.
The results for the intensity threshold method are shown in Fig.~\ref{fig:brightness_threshold}. The best $F_1$ score obtained, of 0.75 for 2$\sigma$, is similar to the best result of the CNN model and surpasses the FNN model. However, if we use a value for the intensity threshold of 5$\sigma$, as suggested by the final recipe given by \cite{2019A&A...626A...4V}, the $F_1$ decreases to 0.2.
The first conclusion from this result is the evident dependence of the threshold value to the observation and a consequence of the method followed to build the ground-truth data for the SDO train, validation and test sets, explained in Sect.~\ref{subsection:SDO_data}. 
This conclusion also explains why CNN models outperform FNNs. CNNs employ a small context window, which helps them to contextualize the properties of the observed pixels, whereas FNN models operate on a pixel-by-pixel basis without utilizing any spatial information.
The second conclusion is that the comparable performance of a non-linear classification technique like CNNs to a thresholding method may suggest that the current approach does not provide significant information for solving the classification problem.

\subsection{Temporal evolution}

A dimension that has not been explored in this work is the temporal evolution of bright patches. If an EB produces a distinct signature in the time domain, this could be used to differentiate it from other bright events and improve the detection. 
Ellerman bombs are dynamic events that evolve in time, producing larger and faster changes in intensity than the typical network bright points. This behavior has been observed in previous studies \citep[e.g.,][]{2000ApJ...544L.157Q, 2007A&A...473..279P, 2011CEAB...35..181H}, where the light curves of EBs detected in $\Ha$ correlated with the temporal evolution of co-spatial brightenings in the mid-UV continua.

\begin{figure}
    \centering
    \includegraphics[width=1\linewidth]{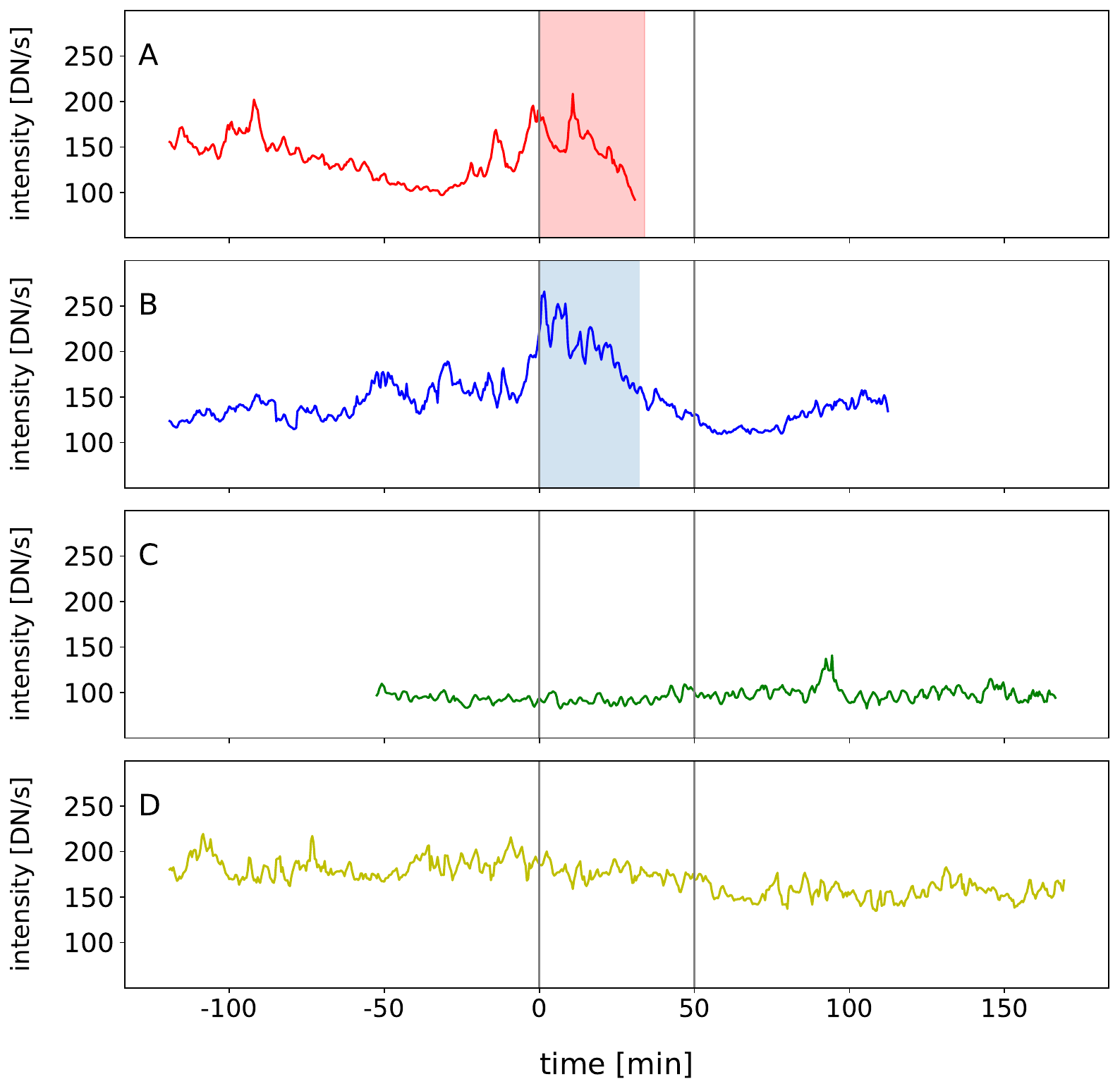} 
    \caption{Temporal variation of the 1600\,\AA\ passband of SDO/AIA bright patches indicated in Fig.~\ref{fig:SDO_thresholds}. Grey lines indicate the interval with SST co-temporal and co-spatial observations. Red and blue shaded areas in the two first panels indicate the EBs detected in $\Ha$ observations associated to bright patches A and B. An offset is applied to the x-axis to have 0 at the start of the SST observation.}
    \label{fig:SDO_lightcurve}
\end{figure}

To test this hypothesis, we analyzed the intensity evolution (or light curve) of the bright patches  indicated with white dashed circles in Fig.~\ref{fig:SDO_thresholds}. We then studied the correlation between their evolution and the occurrence of co-temporal and co-spatial $\Ha$ EBs in the SST datasets.
Figure~\ref{fig:SDO_lightcurve} shows the light curve of these bright patches for the 1600$\,\AA$ passband, extending the temporal domain to include 2~h before and 2~h after the SST observations.
The varying lengths of the light curves are given by the different lifetimes of the bright patches. To compute the intensity at each time, we averaged over 36 pixels around the brightest point of each bright patch.
The two first panels show the light curve for the bright patches A and B, which have an associated EB. The red and blue shaded areas indicate the EBs detected in $\Ha$ observations associated with bright patches A and B. The light curves at 1600$\,\AA$ reveal an important increase in intensity during the EBs' lifetime (about 1.5--2.0 larger in amplitude compared to the minimum intensity at the same location). These EBs are exceptional cases with a very long lifetime, but similar trends also appear in other observations, such as EB-2 in \cite{2015ApJ...812...11V}.
A question arises whether we should consider the entire duration of the event as a single EB, or the concatenation of several EBs at the same location.
In contrast, the two last panels present the light curve for the bright patches C and D, which have no EB associated. Their light curves do not show any strong intensity variations, but the average intensity could be as high as some of the EBs. For example, in panel D,  the intensity is comparable with the intensity reached in the panel A during the associated EB. This indicates that the study of the temporal evolution of AIA bright patches could be the key to detecting EBs effectively using SDO/AIA data.
Thus, the study of the light curve of the bright patches identified by the SDO CNN model can be used as next step to identify EBs with more precision. Indeed, there is a popular type of neural network for spatio-temporal predictions \citep[ConvLSTM;][]{2015arXiv150604214S} which has proven to be a key part to improve the detection of solar far-side active regions by exploiting the temporal information \citep[][]{2022A&A...667A.132B}.

\section{Summary and conclusions}
This study explores the automatic detection of EBs in solar observations using NNs, with the aim of developing a generalizable method applicable to different datasets. The models were trained using high-resolution $\Ha$ observations from the SST and data from the 1600$\,\AA$, 1700$\,\AA$, 171$\,\AA$ and 304$\,\AA$ channels of SDO/AIA.

For the SST observations, NN-based models (both CNN and FNN) proved to be effective in detecting EBs. Spatial information is not crucial for detection when sufficiently high spectral resolution is available. However, as spectral and spatial resolution degrades, spatial information becomes more relevant, as indicated by the CNN models, which incorporate spatial information, and outperform FNN models in these scenarios. The inclusion of $\mu$ did not significantly improve the detection accuracy. Finally, the feature importance analysis revealed that the $\Ha$ line wings at $\pm$1\,\AA\ offset are the most informative for detection.

For the SDO/AIA models, the detection of EBs was more challenging due to the lower spatial resolution and the lack of detailed spectral information. The models failed to adequately distinguish between EBs and bright events of different origin.
This suggests that the combined intensity of the different channels does not provide enough information to classify EBs. The CNN models generally outperformed the FNN models, indicating that spatial information is crucial for detection when the spatial resolution is lower (as indicated by the spatially degraded SST dataset). 

The feature importance analysis showed that the 1600$\,\AA$ and 1700$\,\AA$ channels are the most important for EB detection (with 1600$\,\AA$ being the most relevant of these two). The 171$\,\AA$ and 304$\,\AA$ channels are mainly used to filter out transition-region events or flares with a strong lower-atmosphere component. 

The comparison between the intensity threshold method and the SDO models pointed out the importance of using the context information of each observation to detect EBs. Both the intensity threshold method based on the local background and the SDO/CNN model, which use spatial information, outperform the SDO/FNN model, which does not use any spatial information.

Finally, this study explored the diagnostic potential of the temporal evolution of bright patches to differentiate EBs from other bright events. Preliminary analysis of the light curves suggests that time evolution could be key to detecting EBs. Ellerman bombs exhibit significant intensity variations on short timescales during their lifetime compared to network magnetic concentrations, indicating that incorporating temporal information could improve SDO/AIA detection models.
This result offers a promising avenue towards developing a reliable EB detection method that can be applied to the full SDO/AIA database. For instance, it could be added to the output of the SDO CNN models to further improve the detection. Such a method would, for example, allow analyzing long-term variations in EB occurrence and establishing the impact of small-scale reconnection events in the formation of active regions \citep{2021A&A...647A.188D}.

Models trained to detect EBs in SST observations provide the capability to automatically process large amounts of data in a reliable and precise manner. With the advent of new observatories such as Solar Orbiter \citep[SolO;][]{2020A&A...642A...1M}, the Daniel K. Inouye Solar Telescope \citep[DKIST;][]{2020SoPh..295..172R}, the upcoming European Solar Telescope \citep[EST;][]{2022A&A...666A..21Q}, and the Multi-Slit Solar Explorer \citep[MUSE;][]{2019AGUFMSH33A..07D}, the solar community is embracing the big data paradigm. Methods such as the one presented in this paper will be crucial for properly exploiting the data produced by these new facilities and understanding the contribution of small-scale energetic events to the energy balance of the Sun.

\begin{acknowledgements}
{We would like to thank the anonymous referee for their comments and suggestions.}

This research is supported by the Research Council of Norway, project number 325491, 
through its Centres of Excellence scheme, project number 262622.
This project has received funding from the European Union's Horizon 2020 research and innovation programme under the Marie Skłodowska-Curie grant agreement Nº 945371.
The Swedish 1-m Solar Telescope is operated on the island of La Palma by the institute for Solar Physics of Stockholm University in the Spanish Observatorio del Roque de los Muchachos of the Instituto de Astrofisica de Canarias.
The Swedish 1-m Solar Telescope, SST, is co-funded by the Swedish Research Council as a national research infrastructure (registration number 4.3-2021-00169).
We acknowledge the community effort devoted to the development of the following open-source packages that were used in this work: numpy (\url{numpy.org}), matplotlib (\url{matplotlib.org}), scipy (\url{scipy.org}), astropy (\url{astropy.org}), sunpy (\url{sunpy.org}), pytrack \citep{9927417}, and pytorch \citep{NEURIPS2019_9015}.
This research has made use of NASA's Astrophysics Data System Bibliographic Services.
\end{acknowledgements}

\bibliography{Auto_EBs_SST_SDO.bib}{}
\bibliographystyle{aa}

\appendix
\section{Image patches selection algorithm}\label{appendix:algorithm}
The following algorithm outlines the process for selecting image patch centers to address the data imbalance issue. This algorithm ensures that the selected points adequately represent both EBs and non-EBs, thereby facilitating a more balanced dataset for training the model. To achieve this, the algorithm proceeds through the following steps:

\begin{itemize}
    \item \textbf{Step 1: Define matrix  $W$}. This matrix is a copy of the ground-truth mask, with values of 1 at EB locations and 0 elsewhere. Optionally, this mask can be smoothed. The matrix is then normalized so that the total sum is 1.
    \item \textbf{Step 2: Define value $w$}. Let us assume we have $n$ image patches. The value $w$ represents the sum of the number of 0s and 1s in all the image patches, i.e., pixels event and non-event. Each 0 counts as $-1$ and each 1 as  $+1$. Thus, if there is an equal number of 0s and 1s, $w=0$. Initially, since $n=0$, $w=0$.
    \item \textbf{Step 3: Define the Probability Density Function (PDF)} as $\mathcal{L}=A^{-|W + w|}$. The parameter $A$ is chosen based on the desired sharpness of the probability to select events. In this case, we use $A = e$.
    \item \textbf{Step 4: Iteratively select the image patches.} For each new image patch, $w$ is recomputed and $L$ is updated to select the next point until the desired amount of image patches $N$ is achieved.
\end{itemize}

To avoid image patches repetition, we do not allow the code to use the same image patch centroid twice. We also apply to each new image patch random transformations which include flipping, rotation and transposing.
To clarify these steps, here is the pseudocode for the image patch selection algorithm:

\begin{algorithm}
\caption{\textbf{Image patch} selection algorithm}\label{alg:card_selection}
\begin{algorithmic}
    \State \textbf{Input:} Number of image patches $N$
    \State \textbf{Input:} Probability Density Function (PDF) $W$
    \State Initialize $w=0$
    \State Select random point $x_0$
    \State Initialize number of selected image patches $n = 1$
    \While{$n$ < $N$}
    \State Update $\mathcal{L}$ with $w$
    \State Select next point $x'\sim \mathcal{L}$
    \State Compute new $w$
    \State $n = n + 1$
    \EndWhile
\end{algorithmic}
\end{algorithm}

\begin{figure}[]
\centering
\includegraphics[width=\linewidth]{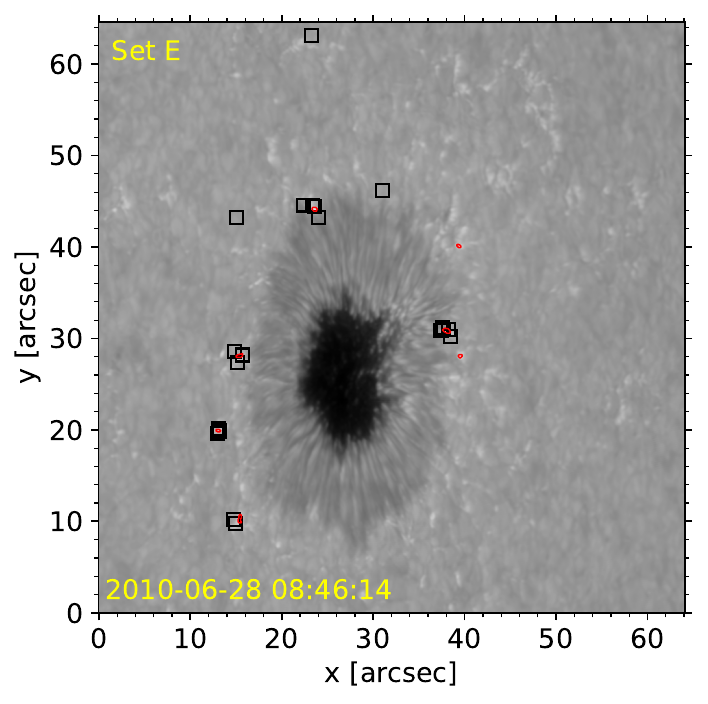}
\caption{Image patches selected by the Image patch selection algorithm in an arbitrary frame of Set E. Image patches are represented by black contour rectangles. Red contours outline the locations of EBs.}
\label{fig:card_selection_algorithm}
\end{figure}

Figure~\ref{fig:card_selection_algorithm} shows the image patches selected for an arbitrary frame of Set E.

\section{Calibration process}\label{calibration_apendix}

The output of the neural network is a value between 0 and 1. This value, as mapped by the network, may not necessarily represent a true probability. Therefore, it is crucial to ensure that these values correspond to actual probabilities. If they do not, we aim to map these outputs into a real probability space.
When the outputs of the model represent true probabilities, the model is said to be calibrated. In this context, a value of 0.7 for a pixel should be interpreted as a 70\% chance of being an EB, and a 30\% chance of being a non-EB.
An uncalibrated model can still be used, but in such cases, the only property that holds is that a higher score indicates a higher probability \citep{fawcett_introduction_2006}. As shown in \cite{DBLP:journals/corr/GuoPSW17}, deeper models tend to show higher levels of miscalibration. Thus, this issue becomes more pronounced with deeper models. Although our models are not particularly deep, we nonetheless address their calibration following the methodology outlined in \cite{DBLP:journals/corr/GuoPSW17}.
It is crucial to note that calibration does not affect the model's capacity but rather improves the interpretability of its predictions.

For the calibration process, we first group the outputs of the neural network into $M$ bins, each one with the same size. In our case, we use intervals defined by the following limits: $[0,0.1,0.2,0.3,0.4,0.5,0.6,0.7,0.8,0.9,1]$. For each bin, we include the lower limit and exclude the upper limit, except for the last bin where both limits are included.

The calibration process is based on two metrics:  \textit{Accuracy} ($acc$) and \textit{Confidence} ($conf$), defined as:

\begin{equation}\label{acc}
acc(B_m) = \frac{1}{|B_m|}\sum_{i\in B_m}1(y_i=1)
\end{equation}
\begin{equation}\label{conf}
conf(B_m) = \frac{1}{|B_m|}\sum_{i\in B_m}\hat{p_i}
\end{equation}
where $B_m $ denotes the $m_{th}$ bin and $|B_m|$ the number of outputs in the given bin. Variables $y_i$ and $p_i$ are the ground-truth label and the associated probability of the output $i$. $Acc$ measures the proportion of correctly classified outputs within a given bin, reflecting the neural network's accuracy. If the third bin ($[0.2,0.3[$) has 100 samples, we would expect that  about 20 to 30 of those samples are correctly identified as EBs. $Conf$ is the average predicted probability within each bin. For a perfect calibration, $acc(B_m) = conf(B_m)$. We can now define the distance between these two quantities as the (average) Expected Calibration Error (ECE):
\begin{equation}
{\rm ECE} = \sum_{m=1}^M\frac{B_m}{n}|acc(B_m) - conf(B_m)|,
\end{equation}\label{ECE}
where $n$ is the total number of outputs. We use this metric to quantify how good is the calibration of a given model. A perfect calibration would have {\rm ECE = 0}. At this point, a non-parametric regression is used to find the mapping function to obtain true probabilities given the output of the neural network, in other words, to find the function which minimizes the ECE. This step is implemented using the {\tt sklearn} python package \citep{pedregosa2011scikit}. After this step, this new mapping is ``attached'' to the neural network as the Calibrator step.

\begin{figure}[]
\centering
\includegraphics[width=\linewidth]{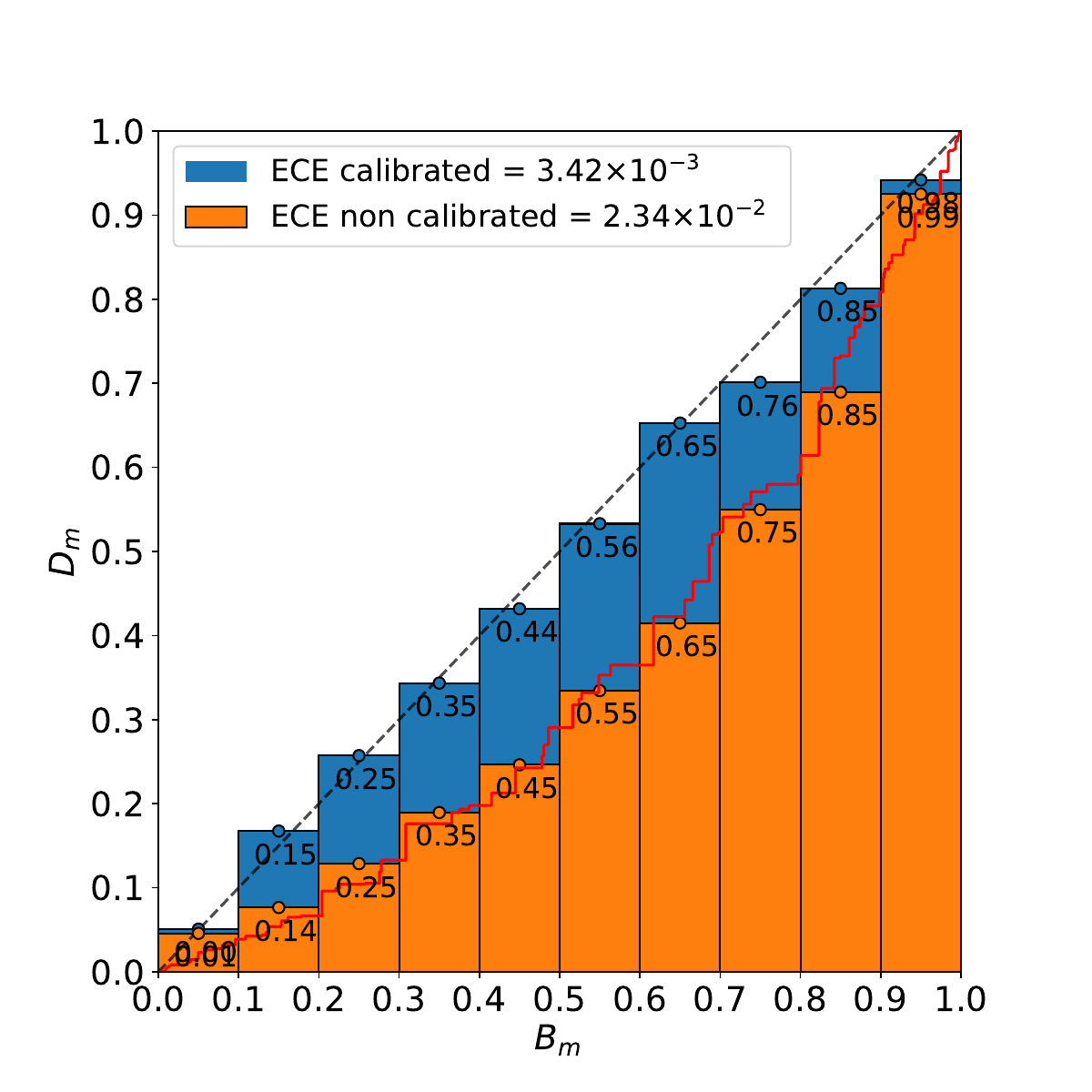}
\caption{Reliability diagram of the CNN model for SST. Orange and blue bars represent the non-calibrated and calibrated model respectively. The number below the head of each bin represents the real $conf$ of that bin. ECE and $\mathrm{ECE\_C}$ are the ECE errors for the non-calibrated and calibrated model respectively. The red curve represents the isotonic fitting.}
\label{fig:calibration_fnn_sst}
\end{figure}

The reliability diagram is a way to visualize the degree of calibration of a model and it compares the deviation of the real confidence w.r.t the theoretical value of each bin defined by:
\begin{equation}
D_m = I_m + (acc(B_m) - conf(B_m)),
\end{equation}
where $I_m$ is the value at the middle of the interval $m$. For example, the third bin comprises the interval [2,3[, so $I_3 = 2.5$. For a perfect calibration, $D_m = I_m$, so the center of each bin should coincide with the diagonal. Figure~\ref{fig:calibration_fnn_sst} displays the reliability diagram for the CNN model of the SST EB identification, before (orange) and after (blue) the calibration procedure, together with the isotonic mapping curve.

\end{document}